\documentclass[msom,nonblindrev]{informs3}
\pdfoutput=1

\usepackage{natbib}
 \bibpunct[, ]{(}{)}{,}{a}{}{,}%
 \def\bibfont{\small}%

\usepackage{pythonhighlight}
\usepackage{balance} 
\usepackage{tabularx}
\usepackage{makecell}

\usepackage{verbatim}
\usepackage{longtable}
\usepackage{cleveref}
\usepackage{multicol}
\usepackage{multirow} 
\usepackage{amsmath}
\usepackage{tabularx} %
\usepackage[noend]{algpseudocode}
\newcolumntype{L}{>{\RaggedRight\hangafter=1\hangindent=0em}X}
\usepackage{rotating}
\usepackage{caption}
\usepackage{threeparttable}
\usepackage{amssymb}
\usepackage{algorithmicx,algorithm}
\usepackage{caption}
\usepackage{amsmath}
\usepackage[T1]{fontenc}

\usepackage{booktabs} 
\TheoremsNumberedThrough     

\EquationsNumberedThrough    

\begin{document}
\TITLE{Agent based modelling for continuously varying supply chains}
\ARTICLEAUTHORS{%
\AUTHOR{Wan Wang$^{1,2}$}
\AFF{$^{1}$School of Transportation and Logistics Engineering, Wuhan University of Technology YuJia Tou Campus No.1178 Heping Avenue Wuchang district, Wuhan City, China \EMAIL{ww1a23@soton.ac.uk}, \URL{}}
\AUTHOR{Haiyan Wang$^{1}$}
\AFF{$^{1}$School of Transportation and Logistics Engineering, Wuhan University of Technology YuJia Tou Campus No.1178 Heping Avenue Wuchang district, Wuhan City, China \EMAIL{hywang777@163.com}, \URL{}}
\AUTHOR{Adam J.Sobey$^{2,3}$}
\AFF{$^{2}$Maritime Engineering, University of Southampton, Southampton, SO17 1BJ, UK $^{3}$ Data-Centric Engineering, The Alan Turing Institute, The British Library, 96 Euston Road, London, NW1 2DB, UK \EMAIL{ajs502@soton.ac.uk}, \URL{}}
} 
\ABSTRACT{%
\textbf{Problem definition:} Supply chains are constantly evolving networks. Reinforcement learning is increasingly proposed as a solution to provide optimal control of these networks. 
\textbf{Academic/practical:} However, learning in continuously varying environments remains a challenge in the reinforcement learning literature.
\textbf{Methodology:} This paper therefore seeks to address whether agents can control varying supply chain problems, transferring learning between environments that require different strategies and avoiding catastrophic forgetting of tasks that have not been seen in a while.  To evaluate this approach, two state-of-the-art Reinforcement Learning (RL) algorithms are compared: an actor-critic learner, Proximal Policy Optimisation (PPO), and a  Recurrent Proximal Policy Optimisation (RPPO), PPO with a Long Short-Term Memory (LSTM) layer, which is showing popularity in online learning environments. 
\textbf{Results:} First these methods are compared on six sets of environments with varying degrees of stochasticity. The results show that more lean strategies adopted in Batch environments are different from those adopted in Stochastic environments with varying products. The methods are also compared on various continuous supply chain scenarios, where the PPO agents are shown to be able to adapt through continuous learning when the tasks are similar but show more volatile performance when changing between the extreme tasks. However, the RPPO, with an ability to remember histories, is able to overcome this to some extent and takes on a more realistic strategy.
\textbf{Managerial implications:} 
Our results provide a new perspective on the continuously varying supply chain, 
the cooperation and coordination of agents are crucial for improving the overall performance in uncertain and semi-continuous non-stationary supply chain environments without the need to retrain the environment as the demand changes.
}


\maketitle

%

\section{Challenges for reinforcement learning in supply chain management}

Automation is increasingly being explored to manage supply chains. Traditionally supply chains have been subject to communication delays and human errors. Retailers worry about the problem of under-stocking which can happen because of an unexpected surge in demand for some products, demand variability, delays, or shortfalls from suppliers/factories.  Stocking inventory in a warehouse is meant to mitigate under-stocking and meet demand, but this is difficult because the supply chain is continuously varying. Analytical approaches have had some success in improving the efficiency of many Supply Chains, however, there are limits to such approaches and they exhibit issues in increasingly complex and uncertain environments \citep{wichmann2020extracting,hosseini2020bayesian}. Therefore, reinforcement learning is increasingly being explored as an alternative for solving complex dynamic inventory problems. For example \cite{chaharsooghi2008reinforcement} found good policies for Supply Chain Order Management under complex scenarios where analytical solutions are not available while \cite{vanvuchelen2020use} demonstrated that Proximal Policy Optimisation (PPO) can tackle the joint order problem, showing that the algorithm approaches the optimal policy.

However, a supply chain is likely to evolve and change over time, from a more lean, stable, and predictable, environment where the variation in demand is low, to something that requires a more agile approach, where the demand is more volatile. This requires an agent to adapt its strategy or for a human user to monitor and retrain the agent.  One of the most appealing qualities of reinforcement learning agents is their ability to continue to learn and improve throughout their lifetime. As such, there has been an increase in interest in continual reinforcement learning \citep{abel2023definition}, a paradigm where agents train on tasks sequentially, while seeking to maintain performance on previously mastered tasks \citep{rolnick2019experience}. While increasing in popularity, there are still limited examples of lifetime reinforcement learning with semi-continuous non-stationarities. The literature points to a number of issues with learning in continuous environments. We need to understand whether agents can transfer learn between tasks, allowing them to rapidly adapt to unseen scenarios and whether they catastrophically forget tasks that they have not seen for some time. To understand these issues recent research in reinforcement learning has developed an understanding of the task capacity \citep{bossens2021lifetime} of an agent, the number of tasks that a policy is able to learn. The number of tasks is partly determined by the interference \citep{kessler2022same} between the tasks, where fewer tasks can be learnt when the tasks are fundamentally incompatible with each other. These concepts are yet to be explored in the Supply Chain management literature, making the determination of whether agents can perform in complex dynamic environments impossible. 

There is an increasing range of Reinforcement Learning approaches being applied to Supply Chain Management, these are highlighted in Table \ref{tab: Characteristics continuous} which summarises the most common Reinforcement Learning algorithms. This literature is split into offline learning, where the agent is trained before the test and doesn’t continue to learn during the operating process, and online learners, which learn as the agent operates. There are some general trends between these different approaches, in particular the growing interest in Recurrent Neural Networks in online learning. However, transfer learning has only been explored in offline learning contexts as the online examples haven't explored evolving environments. 

\begin{table*}[!htb] %
	\centering %
\caption{Characteristics of selected varying continuous supply chain studied in RL} %
	\label{tab: Characteristics continuous} %
	\fontsize{10}{8}\selectfont    %
	\begin{threeparttable} %
		\setlength{\tabcolsep}{0.2mm}{
			\begin{tabular}[]{p{1.2cm}p{1.9cm}p{4.5cm}p{3.1cm}p{1.6cm}p{1.3cm}p{1.7cm}
   } 
				\toprule         
				\multirow{2}{*}{\bf Source }& 
		\multicolumn{6}{c}{\bf Comparison of supply chain methods } \\ 
  \cmidrule{2-7}  
				&Research classification & Reference& Technique & Transfer learning   & RNN &Tests in dynamic environment\\
				\midrule 
    \multirow{25}{1.5cm}{\rotatebox[origin=c]{90}{	 Continuously varying supply chains }}&\multirow{12}{2cm}{Offline learning algorithms}& &   \\
    & & \cite{de2022reward} & DQN & \checkmark &  \textcolor{blue}{X}   &  \textcolor{blue}{X} 
 \\
				& & \cite{oroojlooyjadid2022deep}  & SRDQN &\checkmark & \textcolor{blue}{X}  & \textcolor{blue}{X} \cr
    & & \cite{afridi2020deep}  & DQN &\textcolor{blue}{X} & \textcolor{blue}{X}  & \textcolor{blue}{X} \cr
  &  &\cite{kosasih2021reinforcement}  & Q-learning &   \textcolor{blue}{X} & \textcolor{blue}{X} &  \textcolor{blue}{X}\cr
  & &\cite{meisheri2022scalable} & DQN &\textcolor{blue}{X} &\textcolor{blue}{X}  &\checkmark \cr
   &  & \cite{paine2022behaviorally}  &DQN &    \textcolor{blue}{X}  & \textcolor{blue}{X}  &\textcolor{blue}{X}  \cr
  & &\cite{alves2021applying}  & DDPG,TD3 &  \textcolor{blue}{X}  &  \textcolor{blue}{X}   &\textcolor{blue}{X}   \cr
  & &\cite{gioia2022inventory} & SAC & \textcolor{blue}{X} & \textcolor{blue}{X}   & \textcolor{blue}{X} \cr
   & & \cite{kara2018reinforcement}& Q-learning & \textcolor{blue}{X} &  \textcolor{blue}{X}   & \textcolor{blue}{X}  \cr
       & &\cite{sultana2020reinforcement} &A2C&\checkmark &  \textcolor{blue}{X} &\textcolor{blue}{X}  \cr
		\cmidrule{2-7}
  & \multirow{7}{2cm}{Online learning algorithms} \\
&  & \cite{bermudez2023distributional} & DCPO & \textcolor{blue}{X} & \textcolor{blue}{X}  & \textcolor{blue}{X}    \cr
				& &\cite{peng2019deep}& VPG &  \textcolor{blue}{X}  & \textcolor{blue}{X}  & \textcolor{blue}{X} \cr
    &  &\cite{meisheri2022scalable} &PPO,
A2C & \textcolor{blue}{X} & \textcolor{blue}{X}& \checkmark  \cr
				 & &\cite{stranieri2022deep}& A3C,PPO,VPG&  \textcolor{blue}{X}  & \textcolor{blue}{X}  &\textcolor{blue}{X} \cr
				 & &\cite{kosasih2021reinforcement}& A2C &  \textcolor{blue}{X} & \textcolor{blue}{X}  & \textcolor{blue}{X}\cr

 & &\cite{hubbs2020or}&PPO&\textcolor{blue}{X} &  \textcolor{blue}{X} &  \textcolor{blue}{X} \cr

      & &\cite{mousa2023analysis}&IPPO,MAPPO & \textcolor{blue}{X}  &  LSTM &  \textcolor{blue}{X} \cr
      
    & &\cite{vanvuchelen2020use}&PPO& \textcolor{blue}{X} &  \textcolor{blue}{X} &  \textcolor{blue}{X} \cr
    
 & &\cite{liu2022multi} & HAPPO &\textcolor{blue}{X} &  GRU &\textcolor{blue}{X}  \cr

      & &\cite{gijsbrechts2022can} &A3C&\textcolor{blue}{X}  &  \textcolor{blue}{X} &\textcolor{blue}{X}  \cr
				 & &\cite{bottcher2022solving}&RNN & \textcolor{blue}{X} & LSTM & \checkmark  \cr
     	& &\cite{ park2023adaptive}&SRL-FSA,SRL-PSA & \textcolor{blue}{X} & LSTM & \checkmark  \cr
& &\cite{gijsbrechts2022can}&A3C& \textcolor{blue}{X}  & \textcolor{blue}{X} & \textcolor{blue}{X}  \cr
     & &
     \textbf{Ours}& \textbf{RPPO,PPO} & \checkmark  & LSTM & \checkmark (CRL) \cr

				\bottomrule %
		\end{tabular}}
	\end{threeparttable}
\end{table*}

In offline learning, the experiments focus on whether a network trained on a given set of input observations can perform well as the environment changes. \cite{sultana2020reinforcement} highlight the importance of having models that don’t need additional retraining as new products are constantly introduced and retired. To solve this problem pre-training of the models is explored on different datasets by transferring the learning to supply chains with different numbers of products. This tests the agent’s ability to handle unknown product sets, although it doesn’t require a change in buying strategy. Transfer learning between a production line with 50 objects to one with 70 results in a reduction in performance of 1.8\% for the store on average and 0.7\% for the warehouse, over training on the specific task. This is similar to a transfer learning approach in \cite{oroojlooyjadid2022deep} except that a SRDQN network is explored. DQN networks are demonstrated by \cite{bossens2021lifetime} to have a lower task capacity to PPO in single policy networks, which is more popular in the Supply Chain literature. There are a number of tests including different action sets, cost coefficients, and demand distributions. It is shown that this transfer learning is more efficient than training an agent from scratch, with a reduction in performance of 17.41\% and highlighting the importance in exploring this area. In online learning  \cite{gijsbrechts2022can} doesn’t perform transfer learning but shows that the agent is ambivalent to hyperparameter tuning, indicating that both task capacity and transfer learning would be expected to be high. However, this is only tested on two problems, with changes to warehouse lead time, source lead time, mean demand, and demand variation. Indicating that building on this work towards continuous learning is a promising avenue for research. 

Learning in an online sense allows an agent to deal with the widest possible range of supply chain environments. While there is a growing body of work demonstrating the ability to transfer between tasks, there is no discussion around the task capacity of a learner in a supply chain context or whether it will catastrophically forget a task after some period as continuous learning environments haven't been explored. Therefore, this paper explores continuous learning in supply chain environments. A stochastic environment, \cite{kosasih2021reinforcement} is utilized as a basis for an environment that is extended so that the demand can also be generated in batches, providing a less stochastic and more lean environment. Recurrent networks, through Recurrent Proximal Policy Optimisation (RPPO), are explored as a network type of growing popularity in the literature. Its performance is compared to PPO \citep{schulman2017proximal}, as the leading Reinforcement Learner on a range of problems and which has seen success in Supply Chain Environments while being shown to have the largest task capacity in single policy agents. These two learners are tested across 6 variations of the environment, from the most stochastic to those with the longest batches and their performance is compared. Finally, a range of continuous learning environments is generated, where the environment changes across the 6 environments, to determine how similar the agents find these environments and to allow an understanding of how much agents are able to transfer learn between tasks and avoid catastrophic forgetting.

\section{A pull-based continuously varying supply chain environment}
The environment uses a supply chain scenario developed for exploring Reinforcement Learning, \cite{kosasih2021reinforcement}. It considers a dynamic supply chain that aims to meet customer demands in a continuously varying environment while minimizing costs. In the original environment, the retailer agent faces a stochastic demand only. A batched demand is added to represent a leaner environment, giving a wider range of potential environments that the agent might see and a wider range of tasks for the continuous learning experiments.

\subsection{Mathematical Formulation}
\label{Mathematical Formulation}

\begin{figure*}[h]
	\centering
	
 \includegraphics[width=0.7\linewidth]{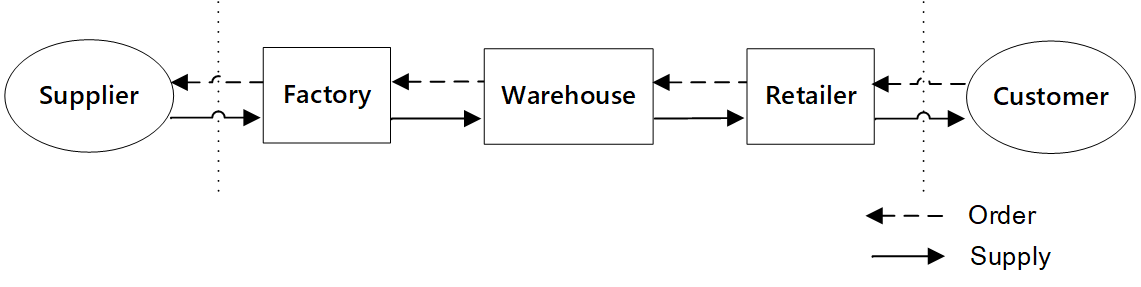}
	\caption{Multi-agent decision-making chain in a continuously varying pull-based supply chain architecture with three agents: retailer, warehouse, and factory. }
	\label{fig:pull-based supply chain}
	
\end{figure*}

The learners in the continuously varying supply chain are arranged sequentially. The customer demand is placed at the retailer, $i=0$, which can initiate the movement of stock by deciding whether or not to place an order. When an order is placed the products move from the factory, $i=2$,  to the retailer. These orders are clipped to conform to the capacity of the environment, for example, the warehouse, $i=1$, can not contain more than 30 items and if more items than this are requested the order is reduced before communicating them.  A similar approach is used to move stock from the warehouse to the factory, this is shown in Figure \ref{fig:pull-based supply chain}. 

The three agents: retailer, warehouse, and factory, work together to maximize the rewards by minimising the sum of the inventory and stock-out costs, shown in Equation \ref{E2}, subject to the constraints, representing the inventory capacity constraints which are 30 for the retailer, warehouse, and factory; 

\begin{equation}
\label{E2}
\begin{aligned}
\max & \sum_{t=0}^{T}  \begin{bmatrix} 
-Hc_{i} \times \sum_{0}^{i}min(I_{i,t},C_{i})\\
-Sc_{0} \times max{ (D_{t}-I_{0,t},0)} \\
\end{bmatrix}\\ \text{subject to:} \\ 
& i \in {0,1,2},\\
&   0 \le I_{i,t} \le C_{i},\\
\end{aligned}
\end{equation}

where $Hc_{i}$ signifies the inventory holding cost for each inventory level, $i$; $I_{i,t}$, represents the inventory at each level at a specific time, t; $Sc_{0}$ represents the stock-out cost, when the retailer is out of stock;  $D_{t}$ represents the demand at the customer, and $C_{i}$ reflects the maximum inventory capacity at each level.

\subsection{Markov Decision Process (MDP) setup}
The Supply Chain Management problem is regarded as a sequential decision-making problem where the ordering decision in the current period affects inventory levels in subsequent periods. The agent aims to maximize its expected cumulative reward in each episode by interacting with its environment in episodes, each consisting of sequences of observations $Obs_{t}$, actions $a_{i,t}$, and rewards $r_{t}$.

\subsubsection{State space}
The state, $s_{i,t}$, is defined as a vector including the inventory level of the retailer, warehouse, and factory, the products in transit between each agent, and the demand for each agent, defined in equation \ref{E8},

\begin{equation}
\label{E8}
s_{i,t}=\left \{I_{i,t}, D_{t} ,O_{i,t}, ST_{i,t} , PT_{i,t}| i\in \left\{ 0,1,2  \right\}, 
\right \} 
\end{equation}

where  $I_{i,t}$, represents the inventory at each level, $D_{t} $ is the customer demand, $ ST_{i,t} $ is the service time  and $ PT_{i,t},i \epsilon \left \{ 1,2\right \} $, is the processing time. $O_{i,t} $ are the pending deliveries in transit between each inventory level.

\subsubsection{Action space}
In each period $t$, agent $i$ observes the new demand, defined in equation \ref{Obs}, 

\begin{equation}
\label{Obs}
Obs_{t} = \{I_{i,t},D_{t},O_{i,t},PT_{i,t},ST_{i,t}|, t\in \left \{ 1,\cdots t  \right \} \}. 
\end{equation}

An action is triggered when the retailer's inventory is below the reorder point and there are no orders in transit; each agent then selects an action, in parallel. The action $a_{i,t}$ is defined as a vector of orders for agent $ i $ and reorder points for the retailer, 

\begin{equation}
\label{E9}
a_{i,t}=\left \{ Q_{i,t}, R_{0,t}  |  i\in \left \{ 0,\dots 2  \right \} \right \},
\end{equation}

where $Q_{i,t}$ represents each agent's order quantity, and $R_{0,t}$, is the reorder point for the retailer, defined by the agent at each time t.

\subsubsection{State transition}
The state transition is deterministic and the transition function is implemented according to the material balance constraints in equation \ref{E6},

\begin{gather}
\label{E6}
\begin{aligned}
I_{i,t+1} &= \min(I_{i,t} + O_{i,t} - O_{i-1,t}, C_{i,t})\\
\text{subject to:}\\
&I_{i,t} + O_{i,t}\le C_{i}\\
&\quad  i \in \{0, 1, 2\}.
\end{aligned}
\end{gather}

This depends on which level the products are transiting between. There is a processing time $PT_{1,t}$, for the warehouse, which remains constant at 3, and $PT_{2,t}$ for the factory, which remains constant at 1, where the demand is held for this number of time steps before release. A service time $ST_{0,t}$ is also defined as 0 at all levels. This means that an order placed will take 4 time steps before it can reach the retailer. The next state $s_{t+1}$ is sampled from a probability distribution $P$ that depends on the entire history, shown in equation \ref{E01},

\begin{equation}
\label{E01}
s_{t+1} \sim P(s_{t+1}|(s_{0},a_{0}),(s_{1},a_{1}),\cdots ,(s_{t},a_{t})),
\end{equation}

where the next state $s_{t+1}$ is determined by the previous state $s_{t}$ and action $a_{t}$. Therefore, the new step inventory $I_{i,t}$ depends on the inventory at the previous step $I_{i,t-1}$, the products received from upstream $O_{i,t}$, minus the products demand downstream $O_{i-1,t}$ .

\subsubsection{Reward function}

If the agent chooses a certain action through trial and error at time step t, then the  $Reward$ can be calculated using equation \ref{E1},

 \begin{equation}
\label{E1}
\begin{split}
\text{$r_{t}$} = & - SC_{0} \times \max{(D_{t}-I_{0},0)} 
\\
& 
- Hc_{0} \times min(I_{0},C_{0}) 
\\
&
- 
Hc_{1} \times 
min(I_{1},C_{1})
\\
&
- Hc_{2} \times min(I_{2},C_{2} ).
\end{split}
\end{equation}

All instances have a per unit inventory cost, which is 1000 for the retailer, 5 for the warehouse, and 1000 for the factory, while the stock-out cost at the retailer is 10000. When an episode is finished, all agents are made aware of the mean episode reward.

The agent will learn to choose the optimal action at each state to maximize its $reward$, minimizing the total supply chain cost $r_{t}$. However, \cite{mousa2023analysis} found that training Reinforcement Learning agents with independent rewards leads to unstable training as well as significantly poorer performance, using shared rewards leads to more stable training as well as better coordination.  In order to minimise the total cost of the supply chain, this paper uses shared rewards to measure the performance of different agents. Notably, our reward function is based on \cite{kosasih2021reinforcement}'s reward function, which increases the retailer's inventory cost. The reward curves in the experiments are the mean episode reward which are averaged over $stats\_window\_size$ episodes, and are 100 by default.
 
\subsubsection{Episode}
The simulation iterations of action $\longrightarrow $ reward $\longrightarrow $ next state $\longrightarrow $ train $\longrightarrow $ repeat, until the end state, is called an episode. An episode ends when the retailer is out-of-stock more than 3 times, or when the number of simulations is greater than 30 days. Each new cycle resets the inventory, with an initial inventory of 10 at the retailer and no stock at the warehouse and factory. The environment follows the structure of an OpenAI gym environment, i.e. the agent interacts with the environment in an identical way (using the "step", "reset" function, i.e. environments must inherit from gym.Env).  

\subsection{ Experimental settings}
The agents were trained using Stable Baselines3 \citep{raffin2021stable,stable-baselines3}. All experiments were run on the IRIDIS5 supercomputer (SLURM, 2023) using CPU Cores Intel(R) Xeon(R) E5-2670, and GPUs (NVIDIA Quadro RTX8000). Different numerical experiments are implemented with different levels of stochasticity. The stochastic environments are taken from \cite{kosasih2022reinforcement} with two ranges of standard deviations which are assumed to require a more agile solution strategy as each product is likely to be different from the last. The batching environments are created to provide a less stochastic environment, with the same range of products as the stochastic environment with a standard deviation of 0.1 but where these products run in product batches, the variables are given in Table \ref{Tab:Test Environments}.

\begin{table}[!htb]\centering
	\caption{Test Environments}
	\label{Tab:Test Environments}
	\begin{tabular}{cccc}\toprule

 \multicolumn{2}{c}{Batch Environment } & \multicolumn{2}{c}{Stochastic Environment}   \tabularnewline
\cmidrule(l){1-2}
\cmidrule(l){3-4}
Name &Range & Name & Range   \\ \midrule
		Bat3 & mean=2, std=0.1,batch\_size=3  &Sto1 & mean=2,std=1\\
  Bat7 & mean=2,std=0.1,batch\_size=7&Sto01 &  mean=2,std=0.1\\
		Bat10 & mean=2,std=0.1,batch\_size=10 & Sto0 &  mean=2, std=0 \\
		\bottomrule
	\end{tabular}
\end{table}

To simulate a stochastic task environment, the customer demand is generated using a normal distribution which is widely adopted in previous studies such as \cite{oroojlooyjadid2022deep,kosasih2022reinforcement,gijsbrechts2022can,liu2022multi} and \cite{park2023adaptive} shown in equation \ref{E18},

\begin{equation}
\label{E18}
f(x) = \frac{1}{\sigma \sqrt{2\pi}} \cdot e^{-\frac{(x - \mu)^2}{2\sigma^2}}
\end{equation}

where $x$ follows a normal distribution with mean $\mu$ and variance $\sigma^{2}$.
Such features simulate customer demand where bursts or stagnant sales randomly occur.

Experiments are also conducted with a batched demand. This allows the representation of supply chains with less variation to a finer fidelity, that might require a more lean strategy allowing a different buying strategy to be developed. The generated batch size can be represented by equation \ref{E19},

\begin{equation}
\label{E19}
X = X \cdot \delta_{ \text{batch\_size}}
\end{equation}

where is the delta function \(\delta_{batch\_size}\) is a positive integer used to specify the number of repetitions and  \(X\) is the original demand random variable following a normal distribution. 

\subsection{Agent  model construction}
\label{PPO and RPPO model construction}
PPO is the base level learner utilized to construct inventory strategies. It shows leading performance on a range of reinforcement learning problems striking a balance between ease of implementation, sample complexity, and ease of tuning trying to compute an update at each step that minimizes the cost function while ensuring the deviation from the previous policy is relatively small. It uses a clipped loss function to provide stable updates, as shown in equation \ref{eq:22},

\begin{equation}
\begin{split}
\label{eq:22}
J^{CLIP}(\theta) = \mathbb{E}\Bigg[&\min\left(\frac{\pi_{\theta}(a|s)}{\pi_{\theta_{\text{old}}}(a|s)}A^{\pi_{\theta_{\text{old}}}}(s, a), \right.\\
&\left.\text{clip}\left(\frac{\pi_{\theta}(a|s)}{\pi_{\theta_{\text{old}}}(a|s)}, 1 - \epsilon, 1 + \epsilon\right)A^{\pi_{\theta_{\text{old}}}}(s, a)\right)\Bigg]
\end{split}
\end{equation}

where $J^{CLIP}(\theta)$ represents the objective function and $\frac{\pi_{\theta}(a|s)}{\pi_{\theta_{\text{old}}}(a|s)}$  is the probability ratio of the previous strategy $\pi_{\theta_{\text{old}}}(a|s)$  and the current strategy $\pi_{\theta}(a|s)$. The probability ratio is then in the range  $[1 - \epsilon, 1 + \epsilon]$ to take into account that the strategy updates should not be too large. It is empirically proven that this results in more stable training behavior.

It is based on an Actor-critic model \citep{konda1999actor}, where the policy improvement in one iteration involves two interrelated neural networks: 

\begin{enumerate}
\item Based on the actor's policy network $pi$, the critic value network $ vf$ learns a policy value function,
\item The actor network estimates the policy gradient and improves its own policy using the critic's policy value function.
\end{enumerate}

A step of critic learning is followed by a step of actor learning. The predicted actions are computed based on the probability of the action policy with a bounded range of probability values. Experience is used to store the agent's experience, specifically, it is a tuple consisting of observation $Obs_{t}$, action $a_{t}$, reward $r_{t}$, new observation $Obs_{t+1}$ and end-of-episode flag done, which is the data required for the agent to learn. By sampling from action experiences and learning through data from different time steps in each episode, these experiences will allow the agent to learn more steadily. An observer is an encapsulation of the environment, $env $, that transforms the state obtained from the environment into a form that is easy for the agent to handle.

\begin{figure*}[!htb]
	\centering
\includegraphics[width=1\linewidth]{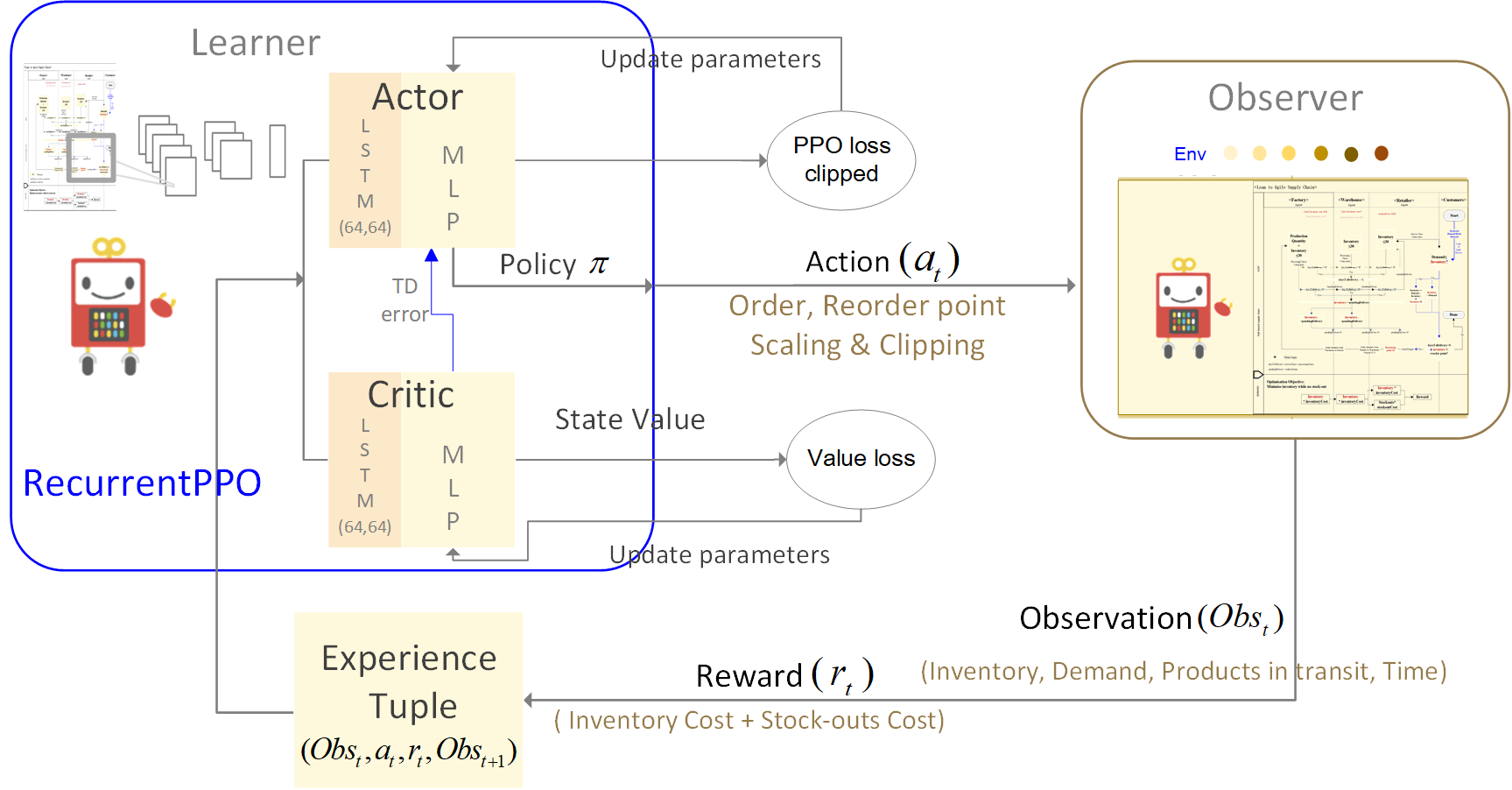}
	\caption{ Agent-environment interaction in Reinforcement Learning using RPPO. The agents learn strategies by simulating actions, interacting with the environment, and obtaining rewards. }
	\label{fig:Agent-environment interaction}
\end{figure*}

Figure \ref{fig:Agent-environment interaction} shows that the learner interacts with the environment in real-time, gathers experience, and updates the policy.

\begin{figure}[!htbp]\centering
	\centering
	\includegraphics[width=0.45\linewidth]{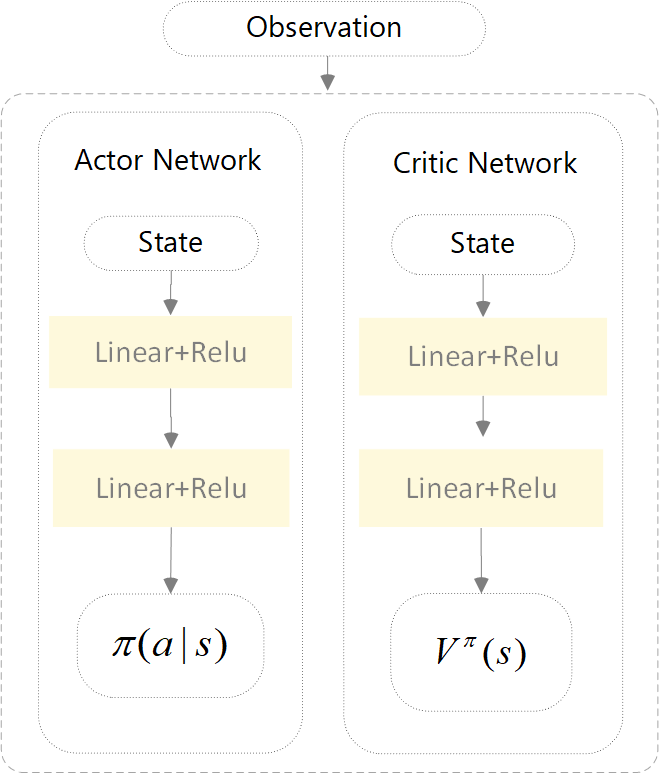} \includegraphics[width=0.45\linewidth]{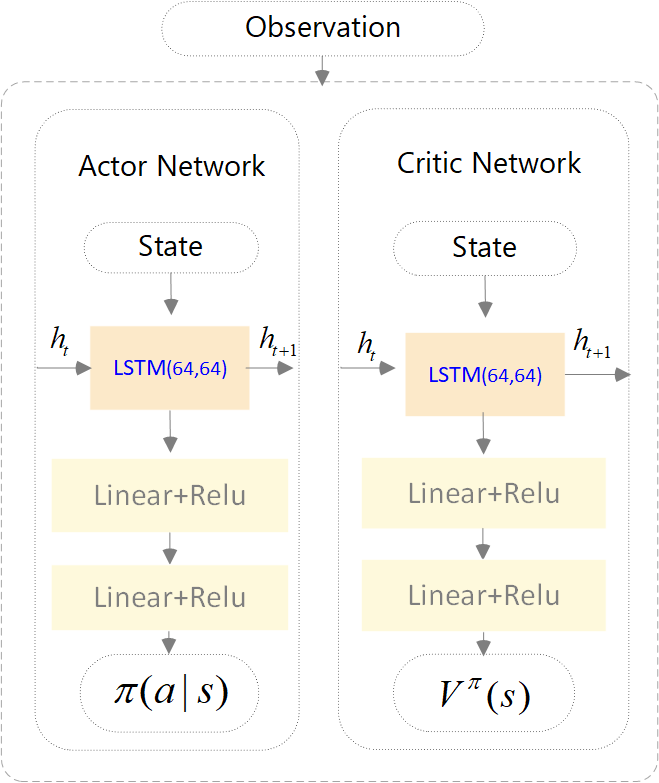}
	\caption{The architecture of learner PPO on the left side and the right side is RPPO with the Recurrent Neural Network.}
	\label{fig: NN architecture}
	
\end{figure}

A recurrent version, RPPO, is also explored where an LSTM network structure, \cite{hochreiter1997lstm}, is added to the the PPO architecture. LSTM is a type of Recurrent Neural Network used to solve the problem of gradient disappearance and gradient explosion during long sequence training. This allows the agent to be able to remember sequences in the past. The general structures of the different networks are shown in Figure \ref{fig: NN architecture}.

Experiments are conducted using the discrete action space on the six different environments, the convergence speed of the RPPO and PPO algorithm are affected by three parameters: the learning rate, $ \alpha  (0 <\alpha  < 1)$, the number of LSTM layers, $ n\_lstm\_layers$, and the number of hidden layers in the critic architecture, $ vf $. Grid search, is a widely adopted hyperparameter tuning technique that systematically explores a set of pre-determined hyperparameter values for a given model and is guided by previous literature indicating the learner's relative stability with respect to these hyperparameters \citep{gijsbrechts2022can}. The final hyperparameters selected based on those that perform best in the grid search are shown Table \ref{Tab:PPO Hyperpara} for the PPO and RPPO agents. 

\begin{table}[!htb]\centering					
	\caption{Hyperparameters used when training the agents. In all settings,these hyperparameters are kept constant.}				
	\label{Tab:PPO Hyperpara}				
	\begin{tabular}{llll}\toprule				
		\textit{Hyperparameters } & \textit{Optimal PPO value}& \textit{Optimal RPPO value}\\ \midrule			
		
  Policy
  & MlpPolicy & MlpLSTMPolicy\\			
  n\_steps &2048&	128\\
  stats\_window\_size	&100&100\\
		Number of epochs &10 &10 \\		normalize\_advantage(bool)&True&	True\\
		Batch size & int = 64 &int=128 \\			
		Learning rate  &  0.003  & 0.003 \\			
        Clip range &  0.2&0.2\\					
         Discount factor (gamma) &  0.99 &0.99 \\					
          Loss &  Mean Squared Error &Mean Squared Error \\					
          Optimizer &  Adam &Adam \\					
          clip\_range &  0.2&0.2\\					
          sde\_sample\_freq &  -1& -1\\					
            entropy\_coeff &  0.0&0.0\\					
            max\_grad\_norm &  0.5&0.5\\					
           gae\_lambda &  0.95&N/A \\					
           vf\_coef &N/A& 0.5\\					
		\bottomrule			
	\end{tabular}				
\end{table}				


\section{Reinforcement learning strategies in different environments}
A series of simulations are run on all six environments: three using varying levels of stochastic demand and three where the demand is grouped into different batch lengths. The policies developed by PPO and RPPO are then compared to an environment with no agent, representing the beginning of the training process. The experiments are repeated 5 times for each environment for PPO and RPPO as the variation in the performance of each agent on each environment is low. 

\begin{figure}[!htbp]
    \centering
    \includegraphics[width=2in, height=0.36\linewidth]{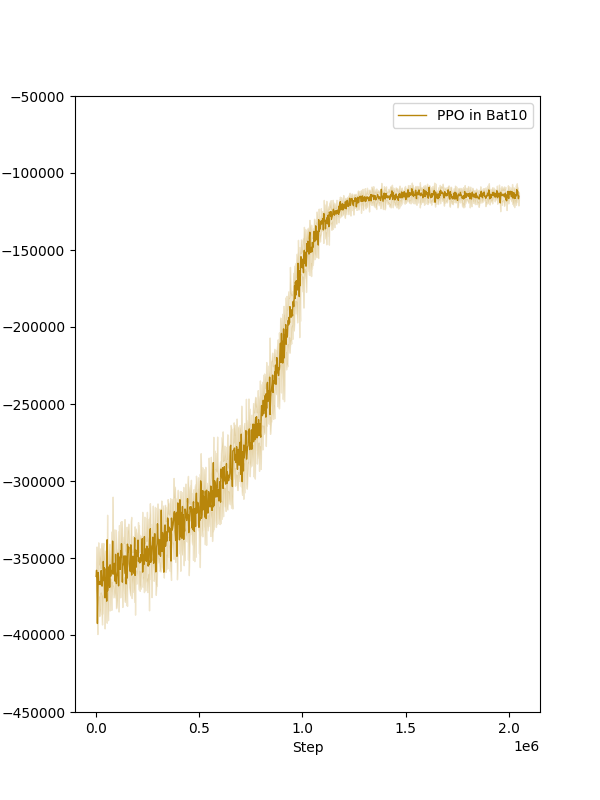}
\includegraphics[width=2in, height=0.36\linewidth]{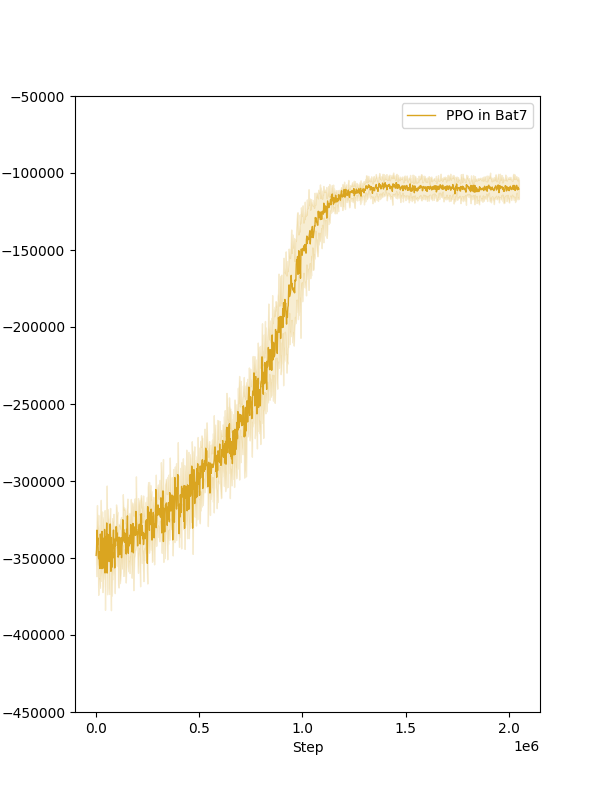} \includegraphics[width=2in, height=0.36\linewidth]{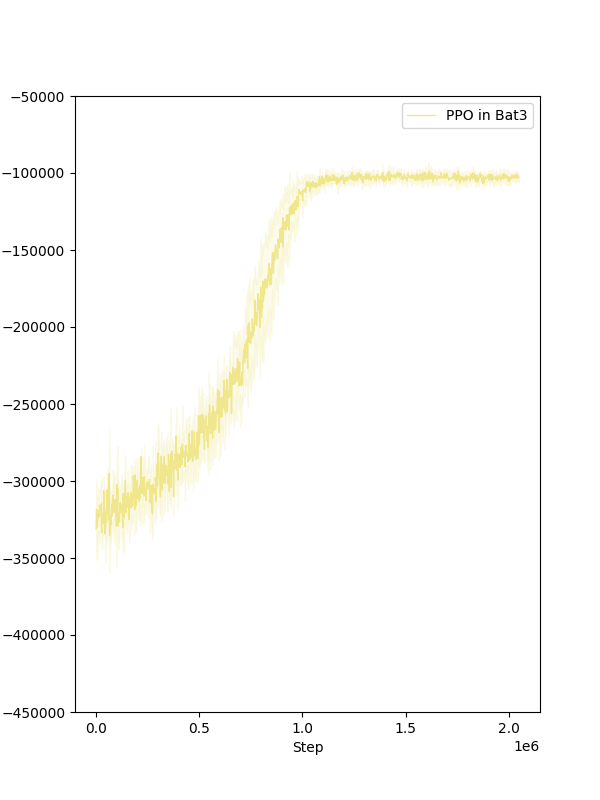}
    \quad    
    \includegraphics[width=2in, height=0.36\linewidth]{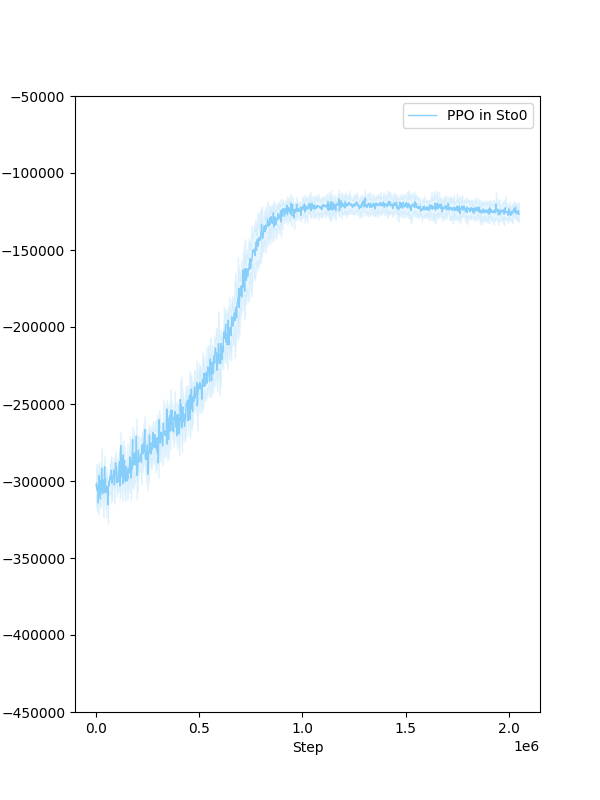}
\includegraphics[width=2in, height=0.36\linewidth]{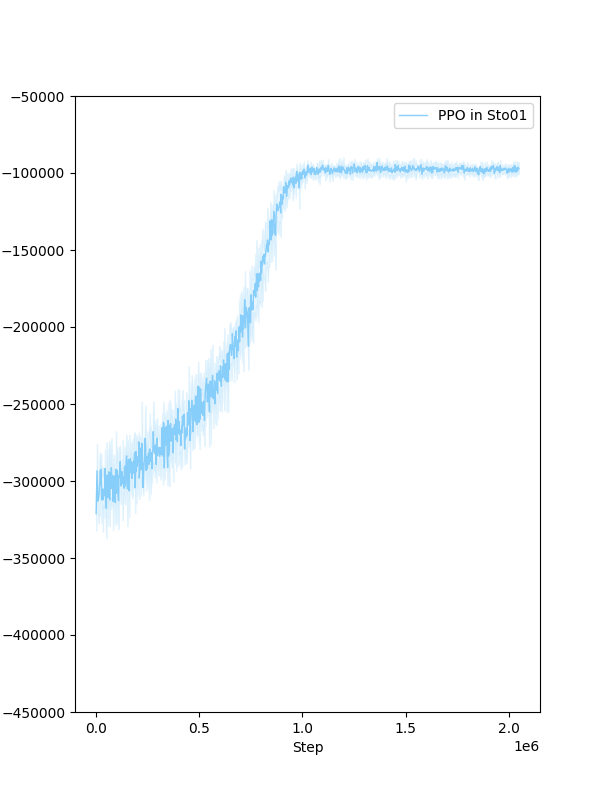}
\includegraphics[width=2in, height=0.36\linewidth]{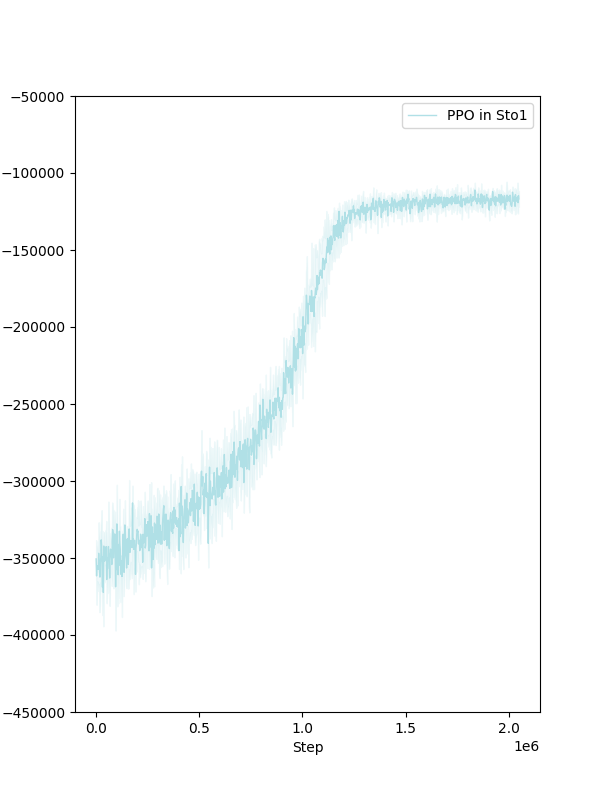}
           \quad 
    
    \caption{PPO agents mean episodic reward curves across tbe six environments.}
    \label{fig:PPO reward curve in six environments}
\end{figure}

The results for PPO are shown in Figure \ref{fig:PPO reward curve in six environments} where the performance across all of the 6 environments is similar, with a final reward of approximately $-1.01 \times 10^5$. For the Batch environments the average reward was $-8.8\times 10^4$ where Batch 7 and Batch 10 converge at approximately $-8 \times 10^4$ with Batch 3 converging a little higher at $-1.02 \times 10^5$. Batch 3 shows the widest range of cycles to convergence and a large standard deviation of 30192, compared with Batch 10 of 8306 and Batch 7 of 3318.

For the Stochastic environments then the performance was similar with the agents converging at a similar value of $-1.15 \times 10^5$ to the learners on the Batch environments but demonstrating a poorer performance. The stochastic learner with no deviation in demand had the worst performance with a mean reward of $-1.35 \times 10^5$ but the agent in the most stochastic environment, 1 deviation, had a reward of $-1.09 \times 10^5$ which was similar to the agent in the next most stochastic environment, 0.1 deviation, $-1.04 \times 10^5$. While the reward curves give similar results to the batch environments, the deviation is higher, 25979 compared to 13938 for the batch environments. With the environment with no variation in demand having the lowest standard deviation of the stochastic environments of 18000.

\begin{figure}[!htbp]
    \centering
\includegraphics[width=1in, height=0.34\linewidth]{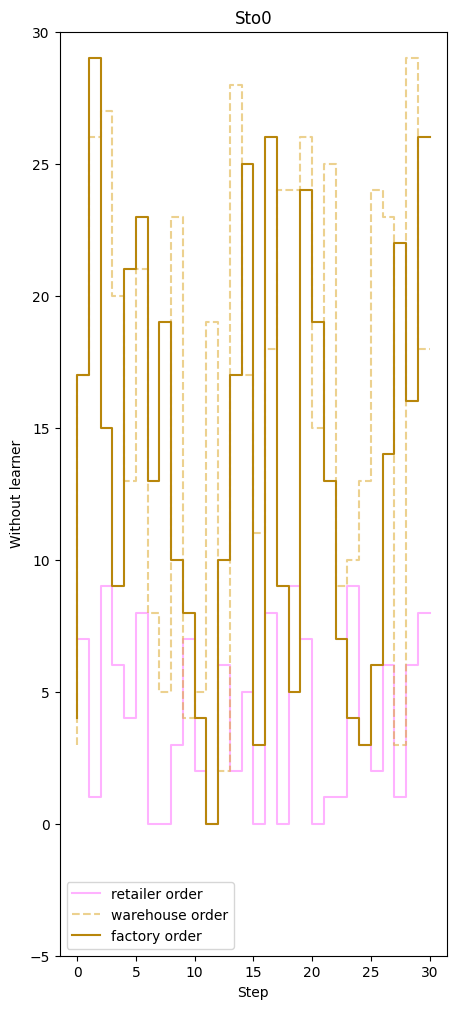}
\includegraphics[width=1in, height=0.34\linewidth]{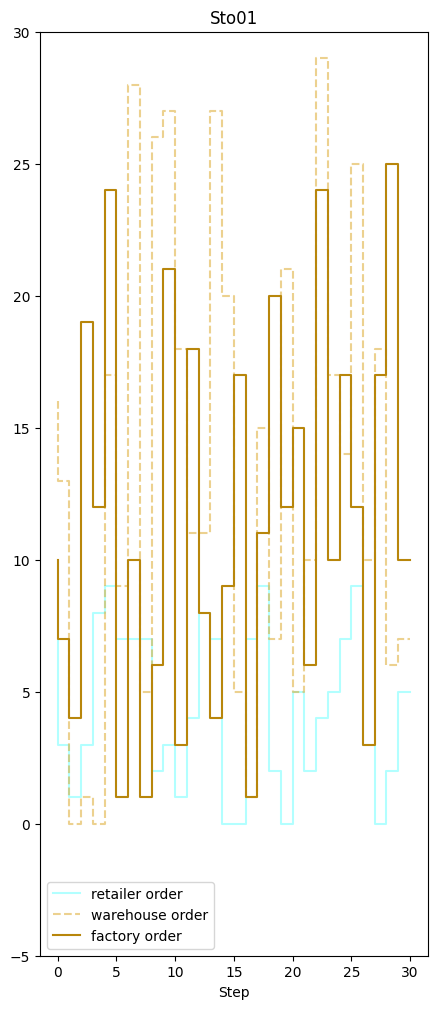}
\includegraphics[width=1.1in, height=0.34\linewidth]{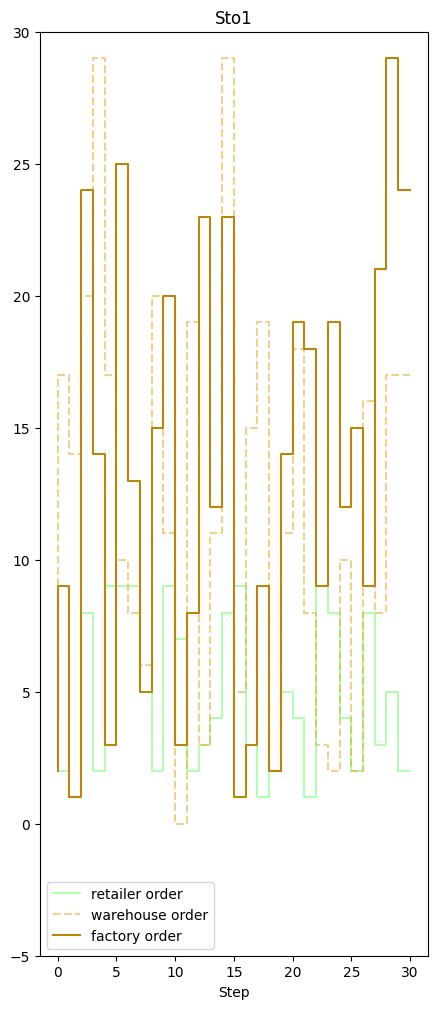}
\includegraphics[width=1.1in, height=0.34\linewidth]{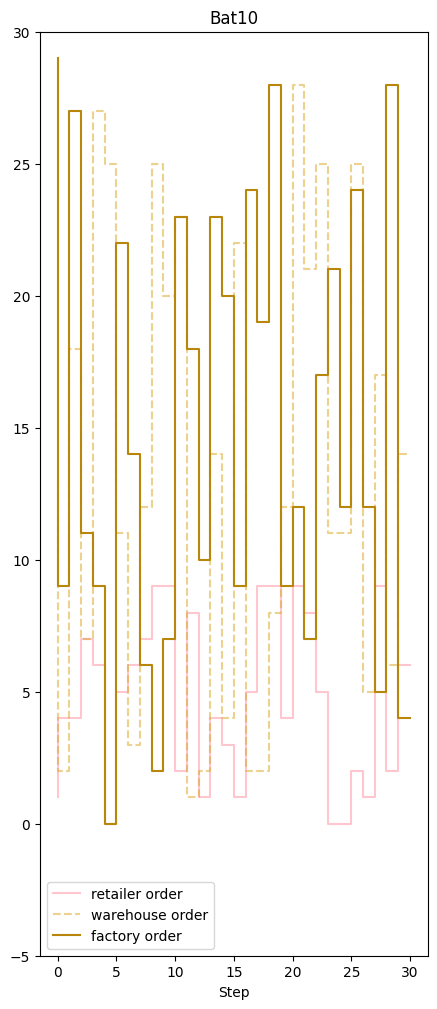}
\includegraphics[width=1in, height=0.34\linewidth]{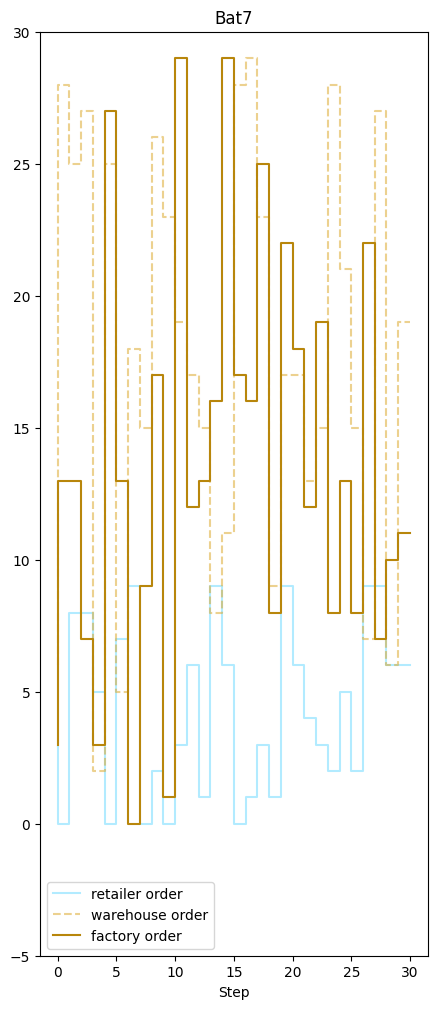}
\includegraphics[width=1in, height=0.34\linewidth]{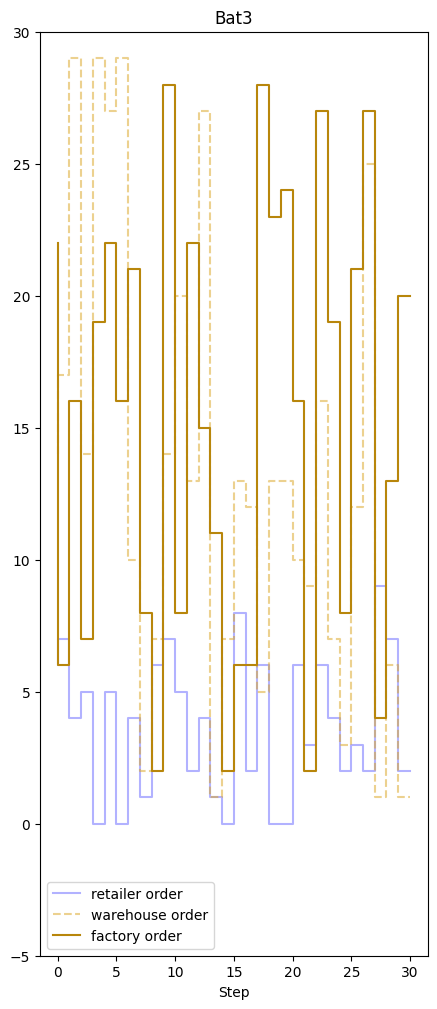}
\caption{No agent order policy visualization, the action spaces for both the warehouse and the factory are discrete values between 0 and 30, while the retailer's is 10. Agents place orders randomly within the action space.}
    \label{fig: random order policy }
\end{figure}

The actions sequences for one episode on each of the different scenarios without a learner are shown in Figures \ref{fig: random order policy }. This illustrates the behaviour at the beginning of the reward sequence where the actions have a high variation, with the warehouse and factory orders varying from close to the maximum of 30 and down to values close to zero. The retailer follows a similar behaviour with a smaller range, represented by it's maximum orders of 10. The behaviours are similar across each of the 6 learners.

\begin{figure}[!htbp]
    \centering
\includegraphics[width=1in, height=0.34\linewidth]{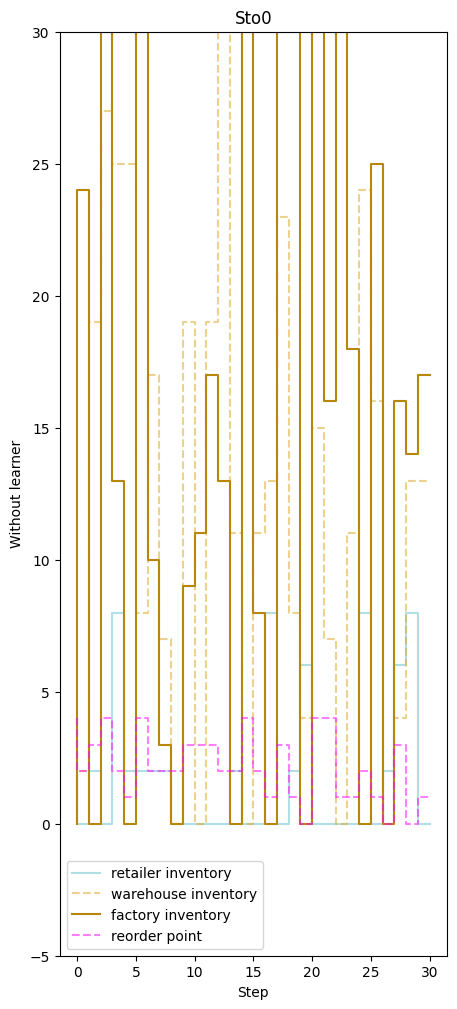}
\includegraphics[width=1in, height=0.34\linewidth]{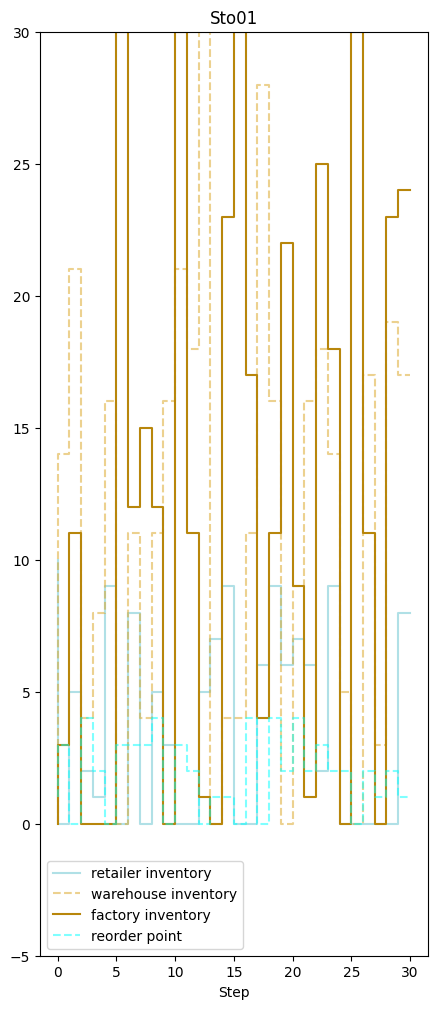}
\includegraphics[width=1.1in, height=0.34\linewidth]{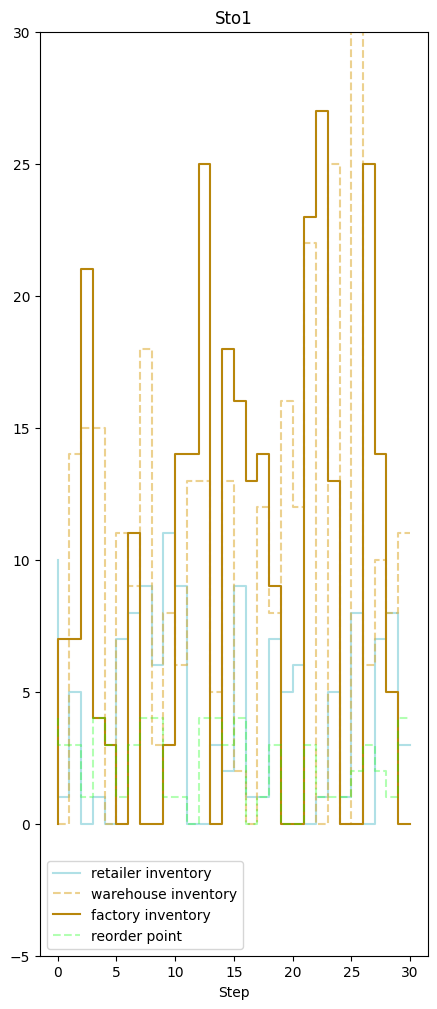}
\includegraphics[width=1.1in, height=0.34\linewidth]{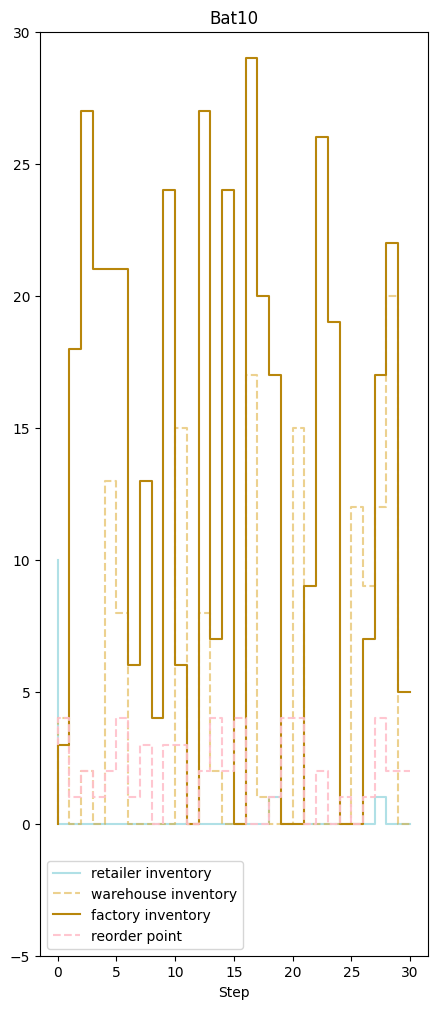}
\includegraphics[width=1in, height=0.34\linewidth]{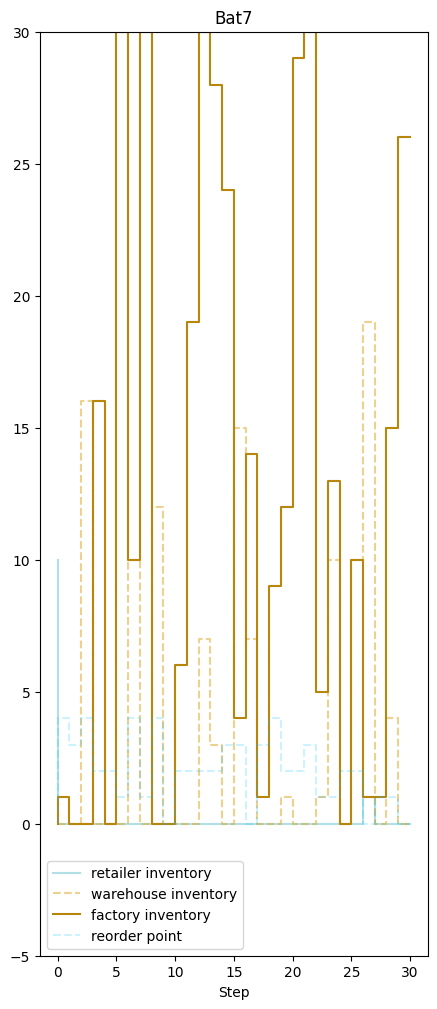}
\includegraphics[width=1in, height=0.34\linewidth]{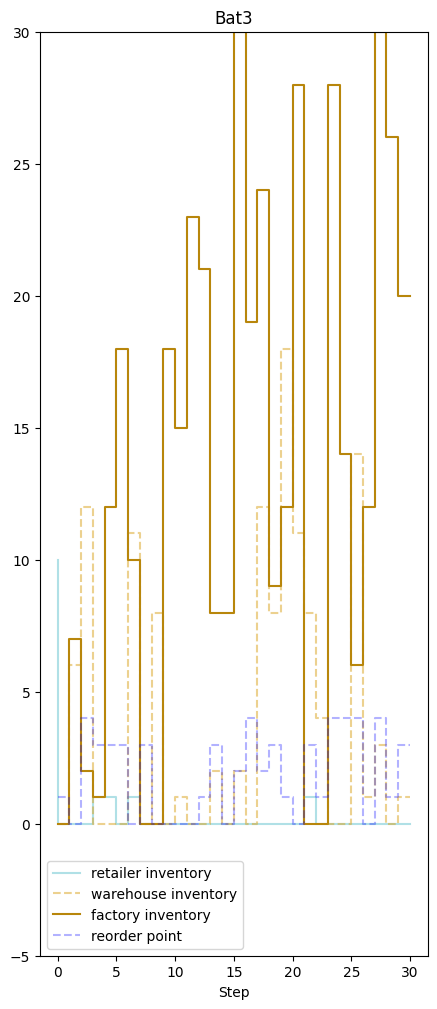}
\caption{No agent inventory levels, representing a random response from the learner}
    \label{fig: random inventory policy }
\end{figure}

The resulting inventory is reflected in figure \ref{fig: random inventory policy }. These inventory levels are often high, at or near the maximum value of 30. The amount of inventory varies from high to low. 

\begin{figure}[!htbp]
    \centering
\includegraphics[width=1in, height=0.34\linewidth]{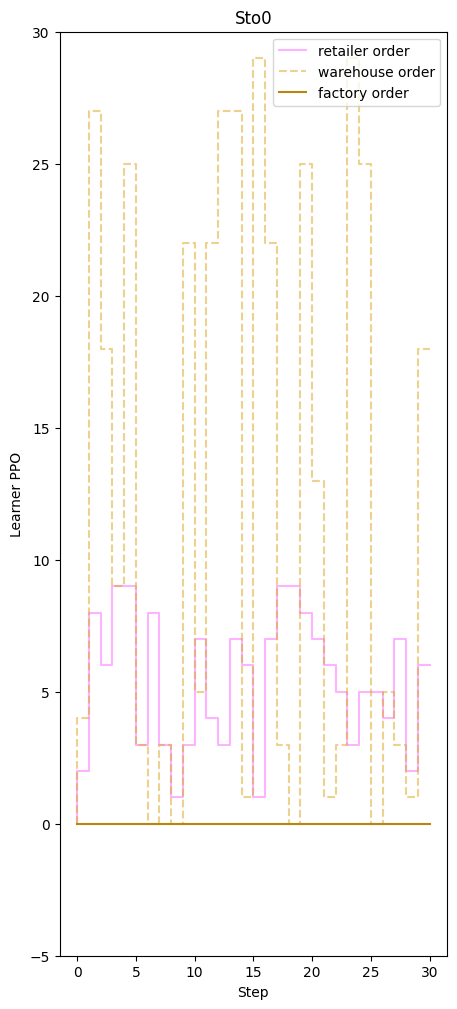}
\includegraphics[width=1in, height=0.34\linewidth]{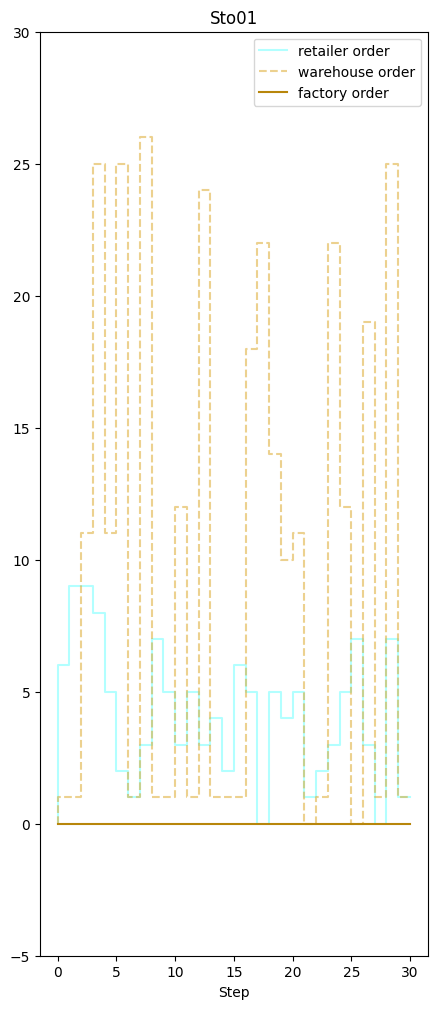}
\includegraphics[width=1.1in, height=0.34\linewidth]{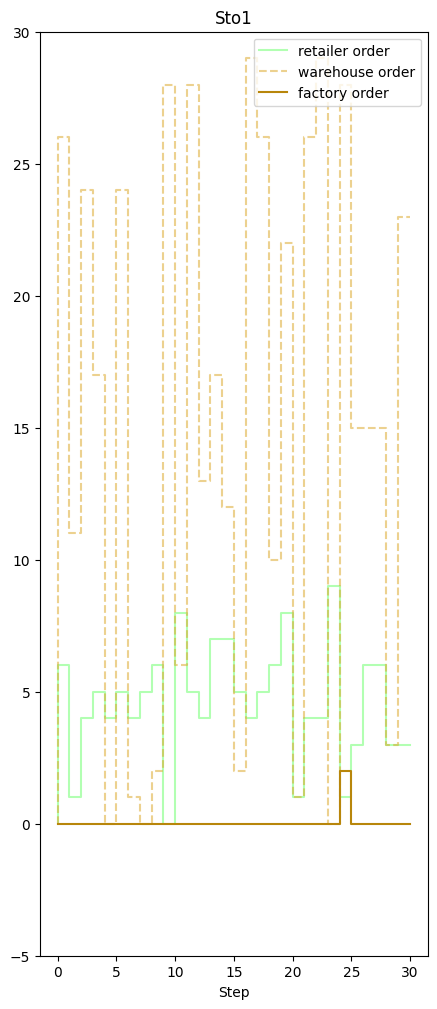}
\includegraphics[width=1.1in, height=0.34\linewidth]{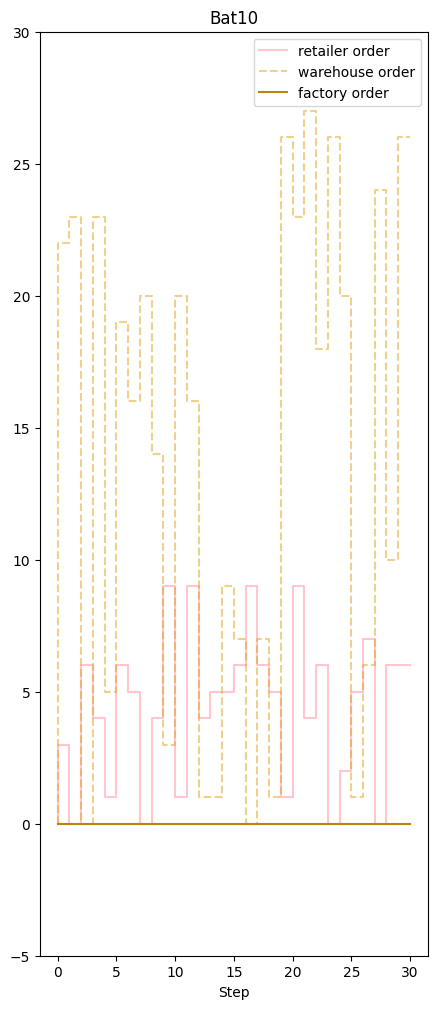}
\includegraphics[width=1in, height=0.34\linewidth]{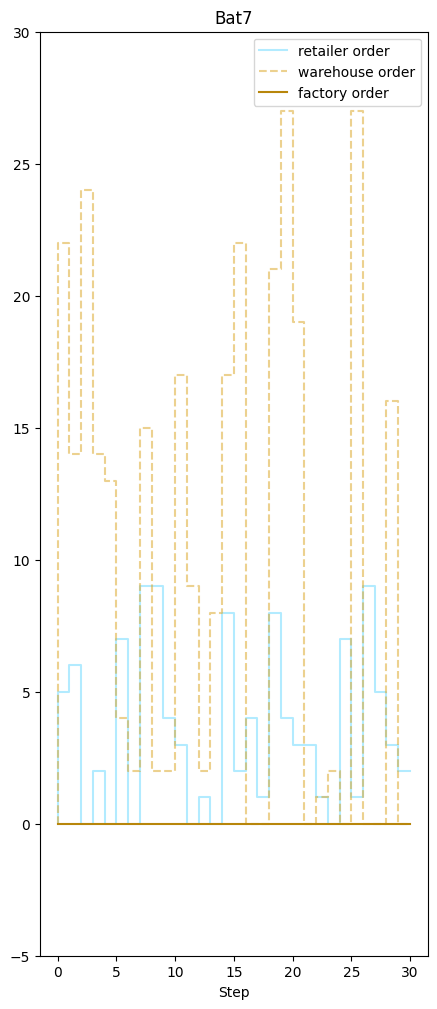}
\includegraphics[width=1in, height=0.34\linewidth]{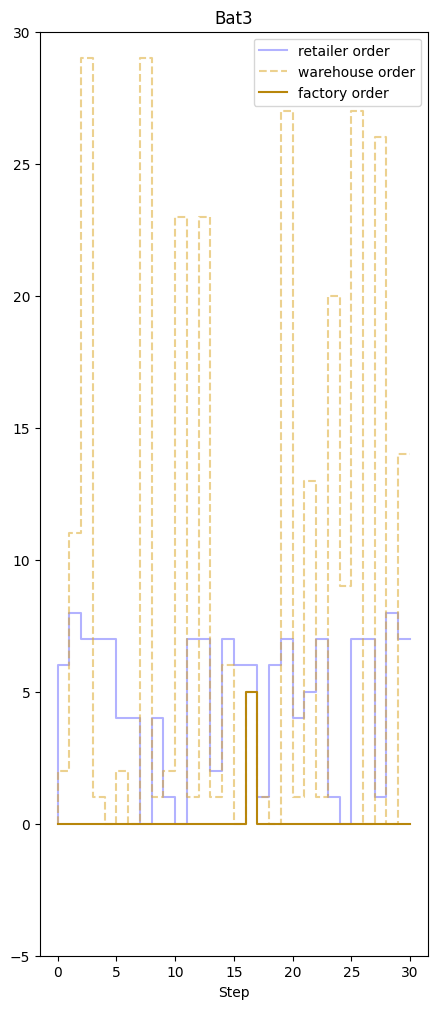}
  \quad    \caption{PPO agents actions, with a limited strategy difference between the stochastic and batch environments }
    \label{fig:PPO order policy }
\end{figure}

The actions for one episode, once the reward cycle for the agent has converged, is plotted to determine whether the PPO agents have formed different strategies. This is plotted as representative but where there is still some difference in performance across the converged state. As the agent gets to a converged state a strategy evolves, as shown in Figure \ref{fig:PPO order policy }. The batch environments and Sto0, which has no variation in the demand, have an average retailer order of 4.04 and the Stochastic Environments have a value of 4.58. All simulations have a maximum of 9 and a minimum of 0 orders, showing that even at this stage that the retailer agent has a high variation. The warehouse agent gives a bigger difference in performance, with the leaner environments showing a lower average order of 9.82 and the Stochastic environments showing a higher average order of 11.52. Again the maximum orders are at 28 or 29 depending on the environment, with a minimum value of 0. However, the factory agent makes no orders and so no stock flows through the system and it will complete early due to stockout. The reorder point action remains at about 2 across all the environments, 2.12 for the less varying environments and 2 for the more stochastic environments.

\begin{figure}[!htbp]
    \centering
\includegraphics[width=1in, height=0.34\linewidth]{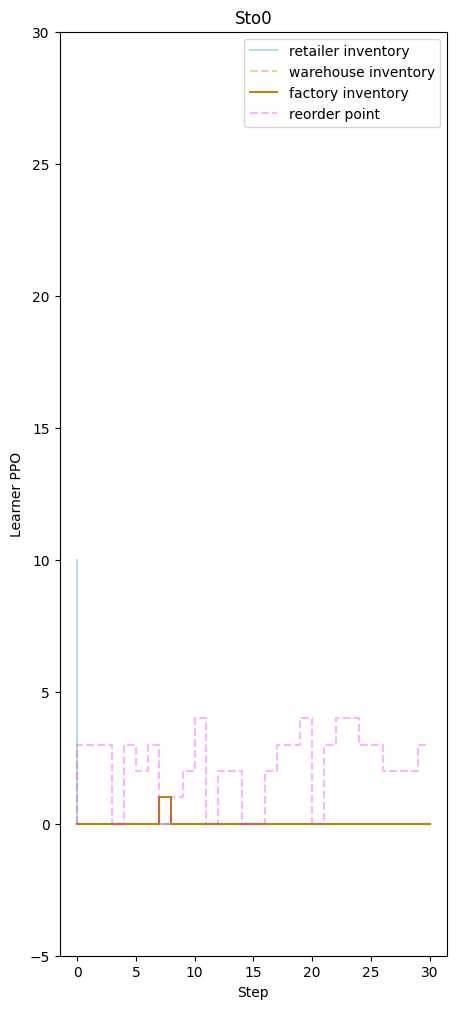}
\includegraphics[width=1in, height=0.34\linewidth]{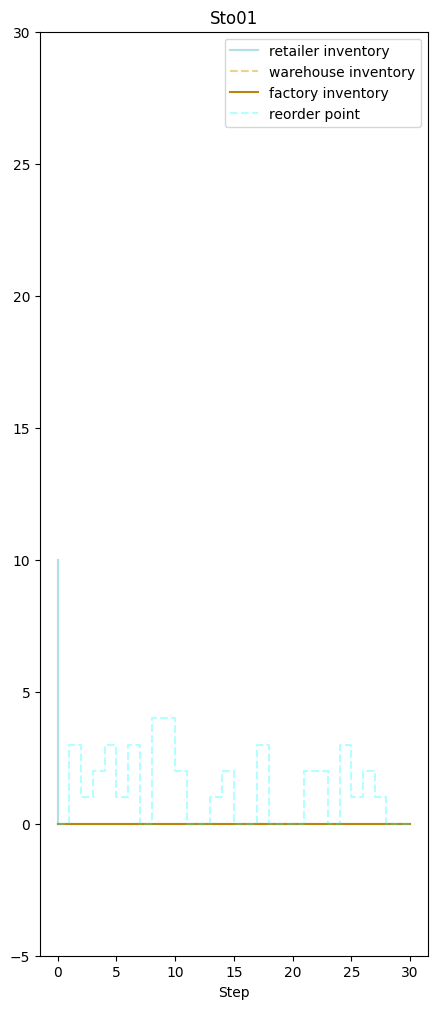}
\includegraphics[width=1.1in, height=0.34\linewidth]{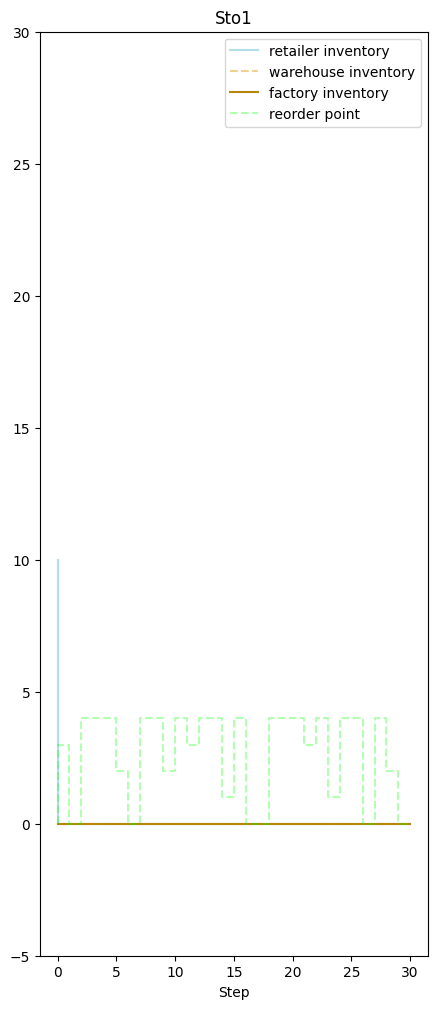}
\includegraphics[width=1.1in, height=0.34\linewidth]{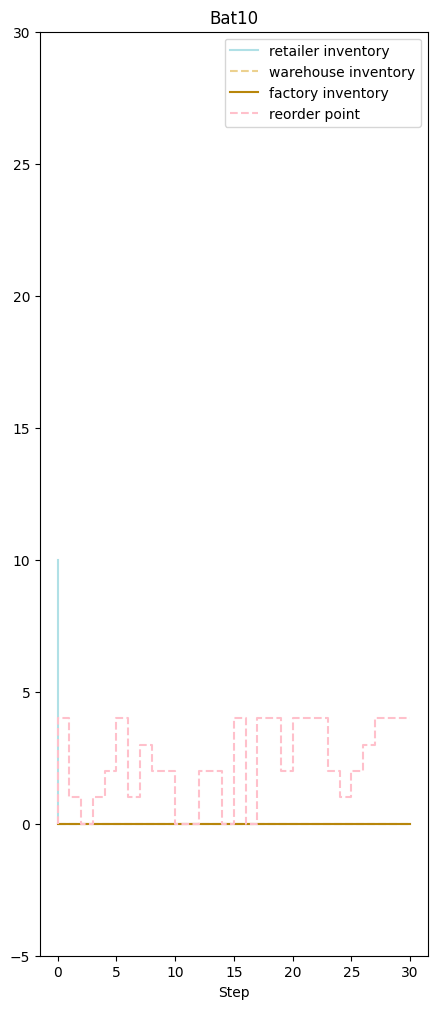}
\includegraphics[width=1in, height=0.34\linewidth]{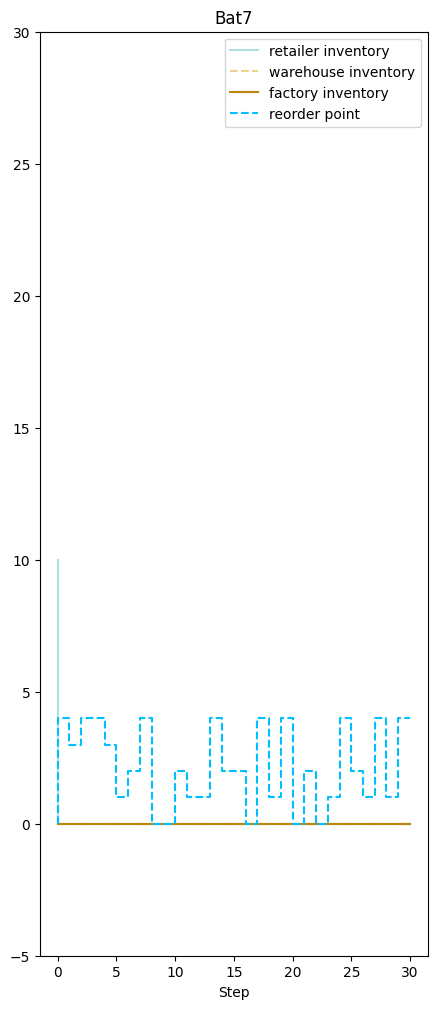}
\includegraphics[width=1in, height=0.34\linewidth]{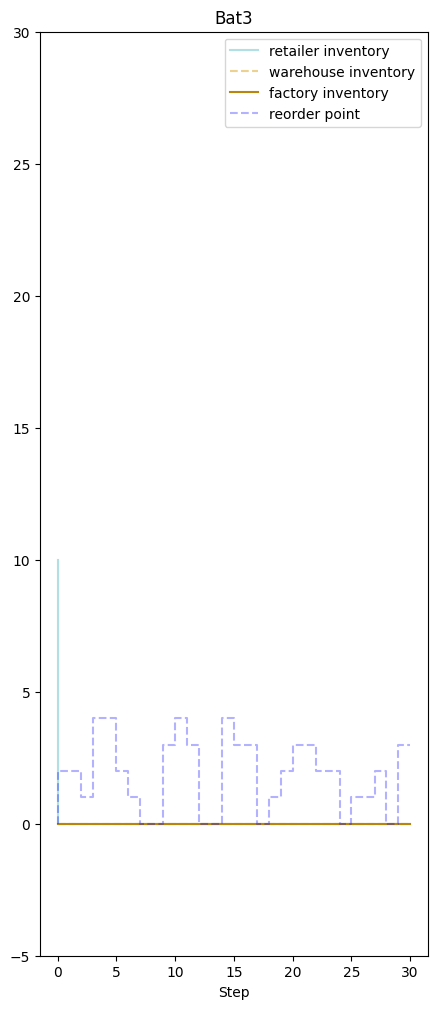}
  \quad    \caption{ Resulting inventory levels showing a limited difference in strategy from PPO agents on the batch and stochastic environments}
    \label{fig:PPO inventory policy }
\end{figure}

Both strategies result in a low amount of inventory in the warehouse and the factory, which is shown in \ref{fig:PPO inventory policy }. For the retailer inventory also stays low. with a value of 0.32 for the batch and no variation environments and a value of 0.42 for the more variable environments. Showing a small difference in performance, although this varies from 10, at the initial state, to 0 as the inventory empties.

\begin{figure}[!htbp]
    \centering
\includegraphics[width=2in, height=0.36\linewidth]{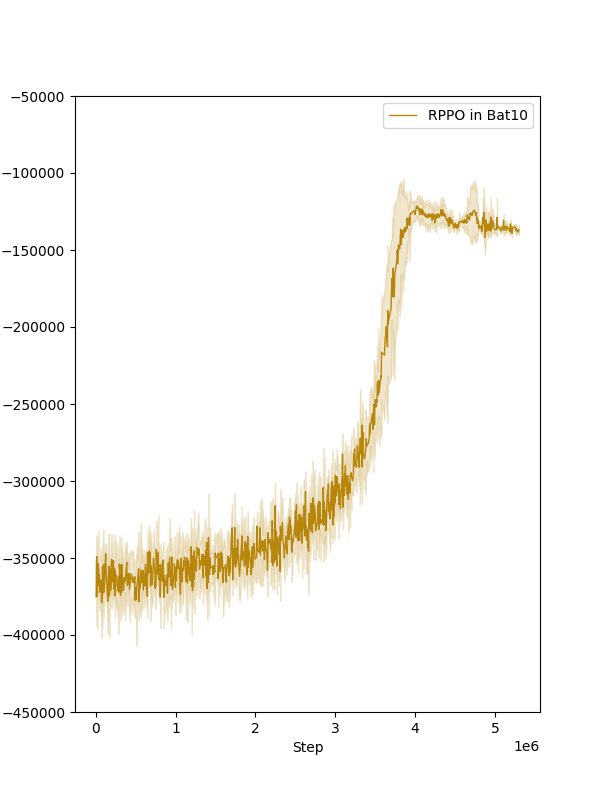}
\includegraphics[width=2in, height=0.36\linewidth]{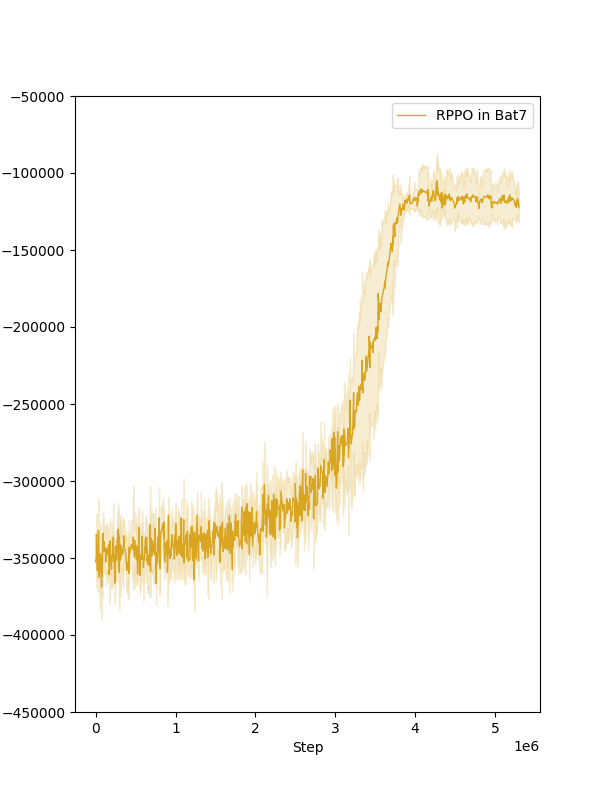}
\includegraphics[width=2in, height=0.36\linewidth]{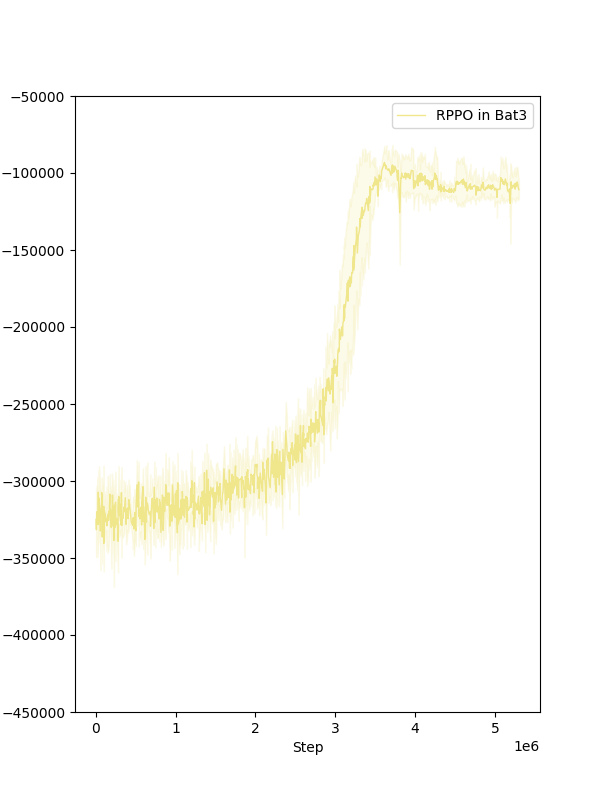}
    \quad 
\includegraphics[width=2in, height=0.36\linewidth]{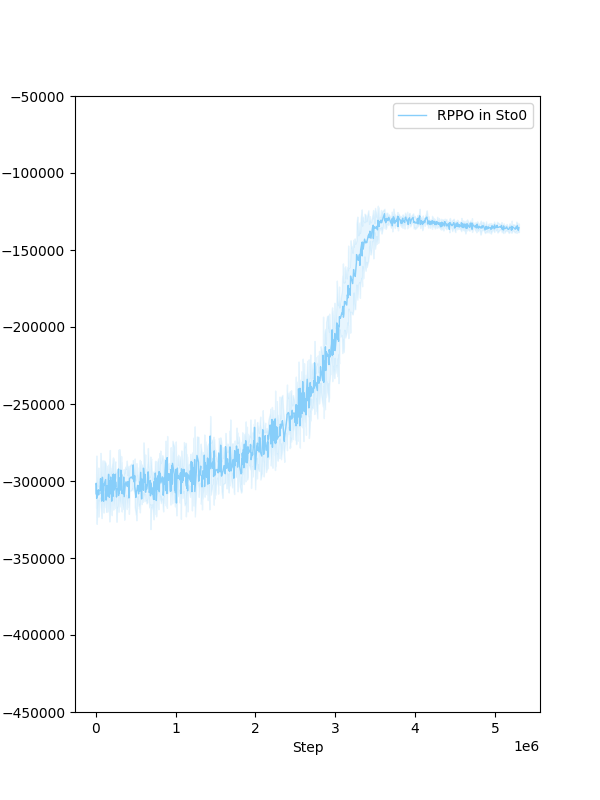}
\includegraphics[width=2in, height=0.36\linewidth]{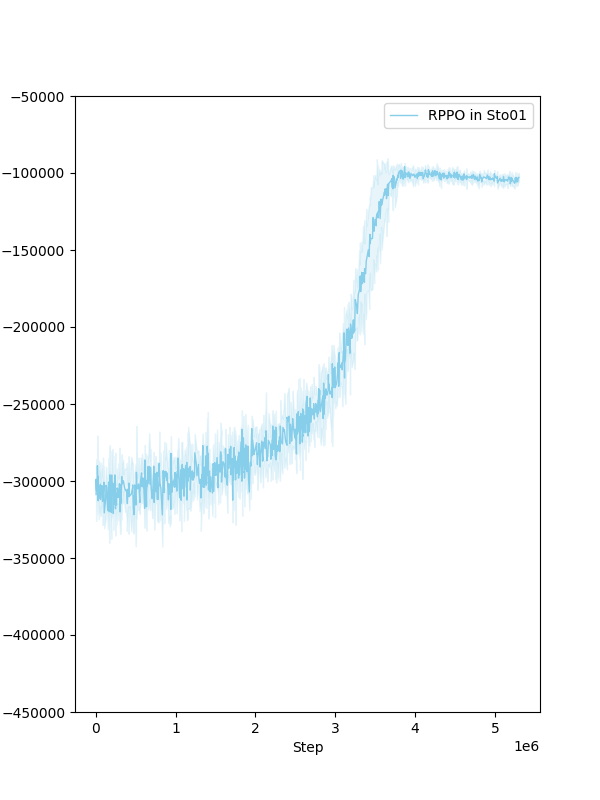}
\includegraphics[width=2in, height=0.36\linewidth]{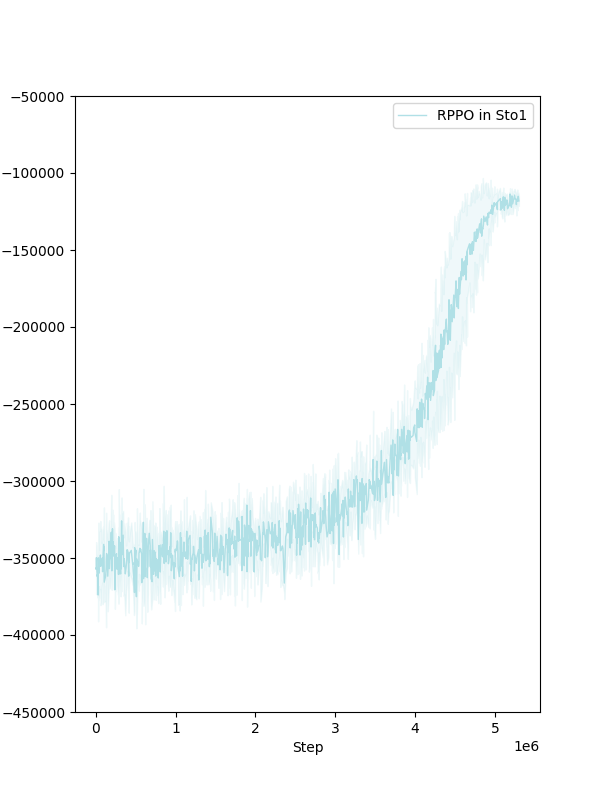}
    \caption{RPPO agents mean episodic reward curves across the six environments.}
    \label{fig:RPPO reward curve in six environments }
\end{figure}

Figure \ref{fig:RPPO reward curve in six environments } shows the training curves on the same six environments for RPPO. In this case then the agents converge at a similar value to those of the PPO learners, of about $-1.09 \times 10^5$ this varies from $-1.32 \times 10^{5}$ in the Batch 10 environment and $-7.3\times 10^{4}$ in the Stochastic 0 environment. However, the RPPO learners require substantially longer training times to reach this performance with convergence at an average of $-4 \times 10^6$ cycles compared to $-1.1 \times 10^6$ for the PPO agents. There is also a difference in the variation of the converged learners, with the RPPO showing a similar variation in the stochastic environments with variation, 33648 for the stochastic environments compared to 29968 for PPO. The same occurs for the batch and zero variation environment with standard deviations on average of 14954 for PPO and 18061 for RPPO. However, the environment with no variation of demand shows no standard deviation in final reward, while the Batch environments show  a higher value of 24082.

\begin{figure}[!htbp]
    \centering
\includegraphics[width=1in, height=0.34\linewidth]{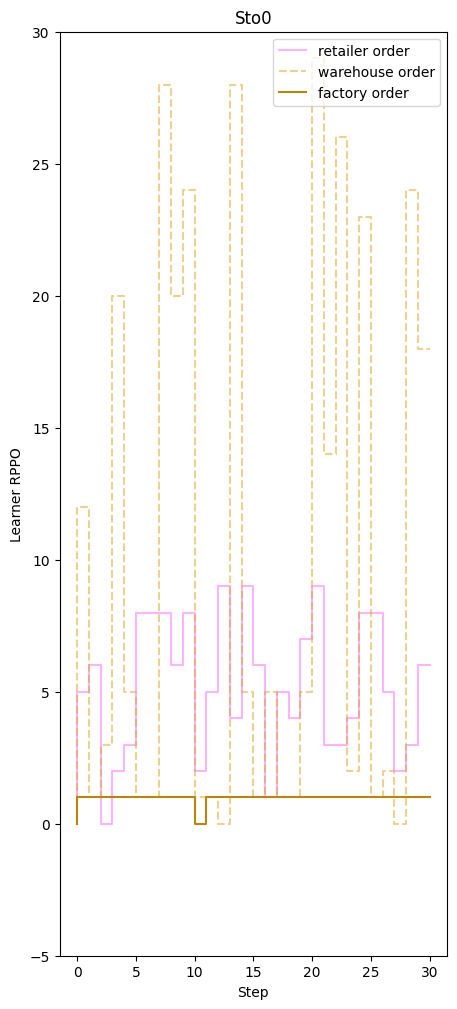}
\includegraphics[width=1in, height=0.34\linewidth]{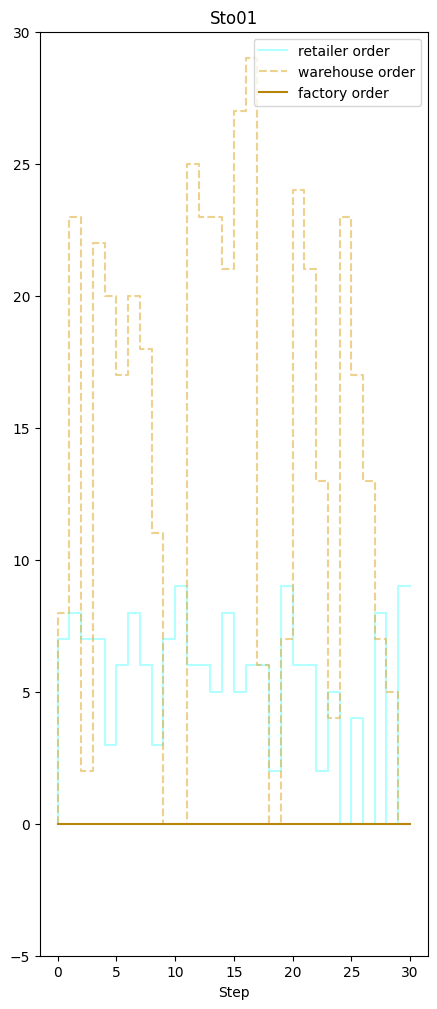}
\includegraphics[width=1.1in, height=0.34\linewidth]{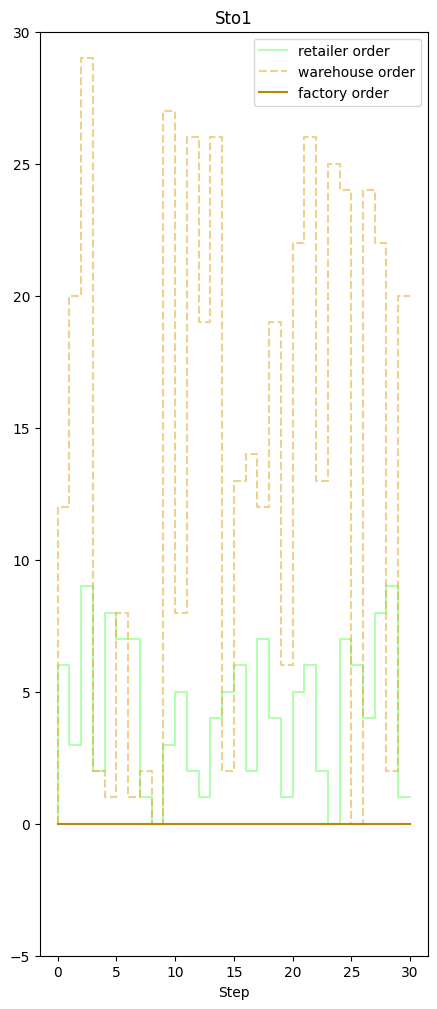}
\includegraphics[width=1.1in, height=0.34\linewidth]{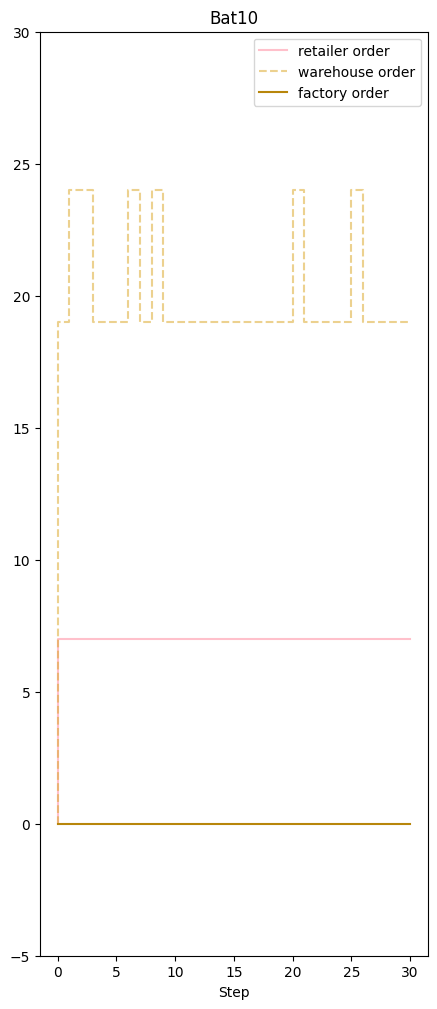}
\includegraphics[width=1in, height=0.34\linewidth]{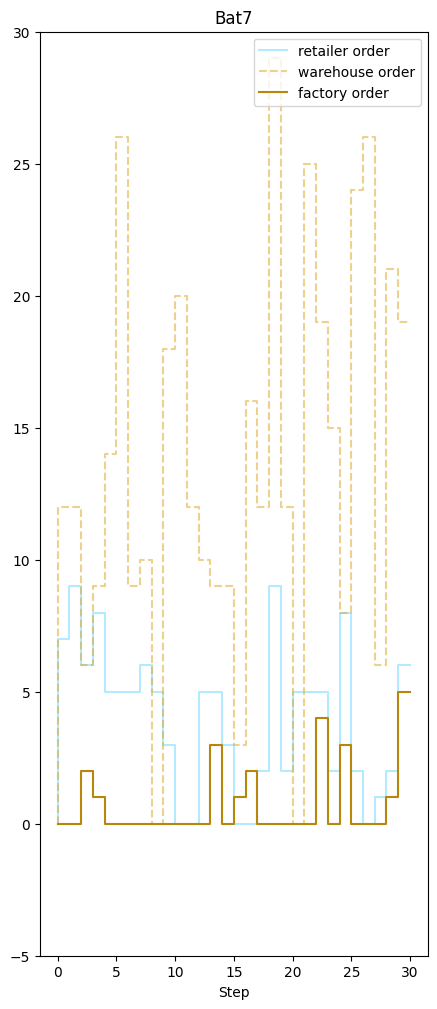}
\includegraphics[width=1in, height=0.34\linewidth]{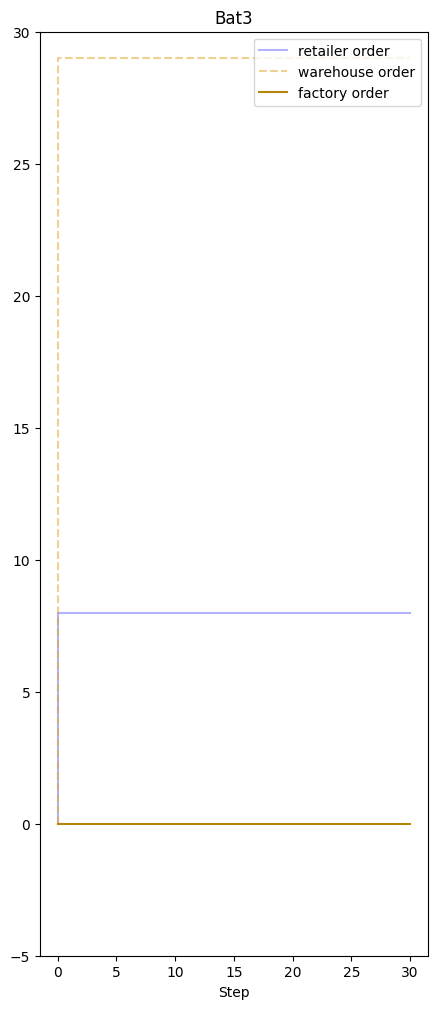}
  \quad
\caption{RPPO agent actions with a different strategy between the stochastic and batch environments}
    \label{fig:RPPO order policy }
\end{figure}

The action sequences for the RPPO agents are shown in Figure \ref{fig:RPPO order policy }, there is a clearer separation in strategies between the two different types of environments. In this case the retailer actions are higher for the lower varying environments, 5.75, with the agent aiming to better predict the frequency of demands but with the higher varying environments the agent has a lower action score. This is also reflected in the warehouse action scores, where the lower varying environments have an action of 17.86 but the lower varying environments show an average order of 12.84. The factory agent behaves in the same way as in the PPO experiment, hardly making any orders. This is slightly higher in the Batch 7, 1.74, and Sto0, 0.935, environments, meaning that in the most lean environments the agents are able to work together to keep the inventories lower through an ability to better predict the future behaviour. This means that these agents are more likely to avoid the stockout loophole and sees the agents trying to generate a low amount of orders, indicating that these problems are easier to solve. The reorder point is different to the PPO and between environments, with a value of 1.58 for the more stochastic environments and 0.98 for the less varying environments. This means a little and often approach to the ordering, making it easier to control the stock as there are no benefits to bulk ordering in this environment.

\begin{figure}[!htbp]
    \centering
\includegraphics[width=1in, height=0.34\linewidth]{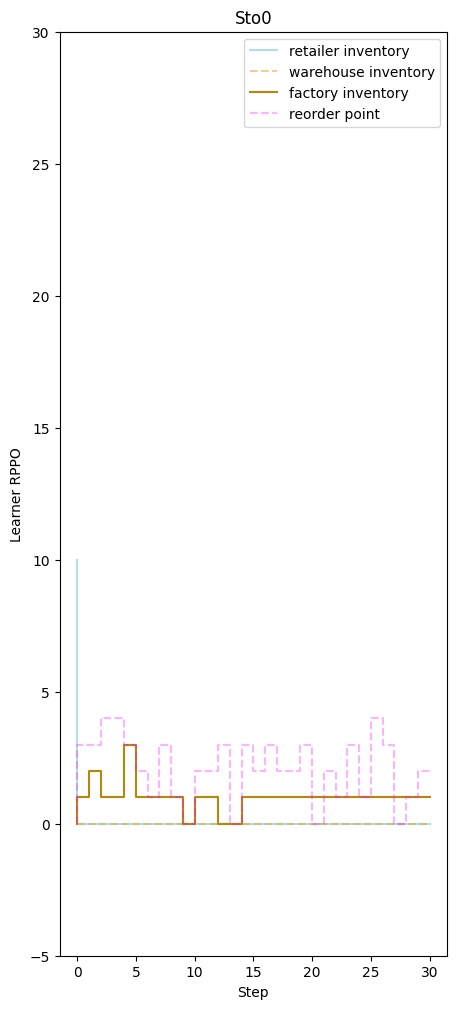}
\includegraphics[width=1in, height=0.34\linewidth]{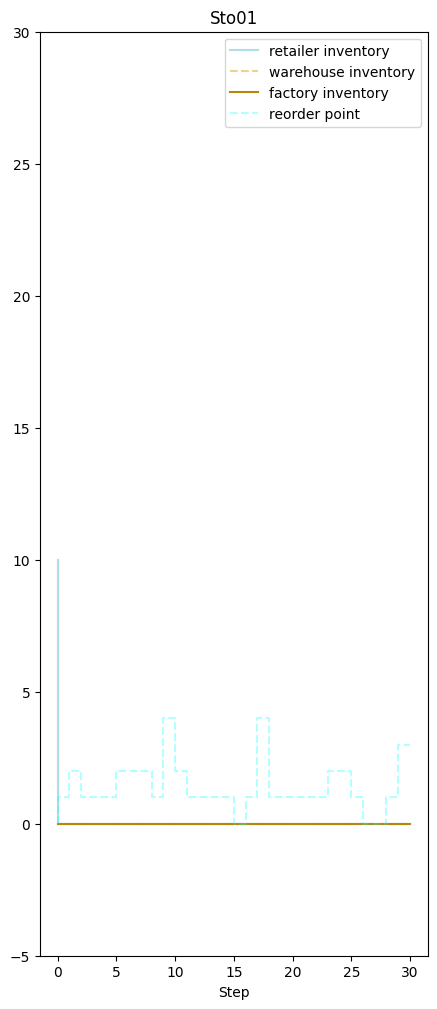}
\includegraphics[width=1.1in, height=0.34\linewidth]{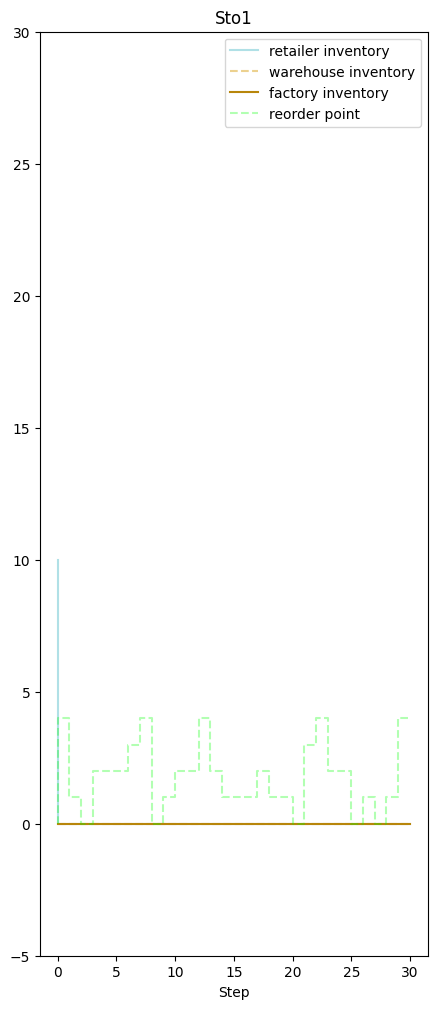}
\includegraphics[width=1.1in, height=0.34\linewidth]{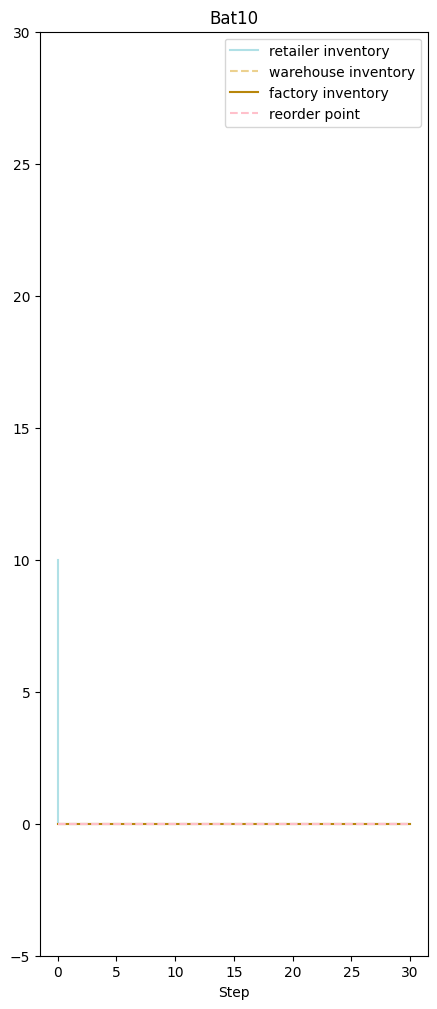}
\includegraphics[width=1in, height=0.34\linewidth]{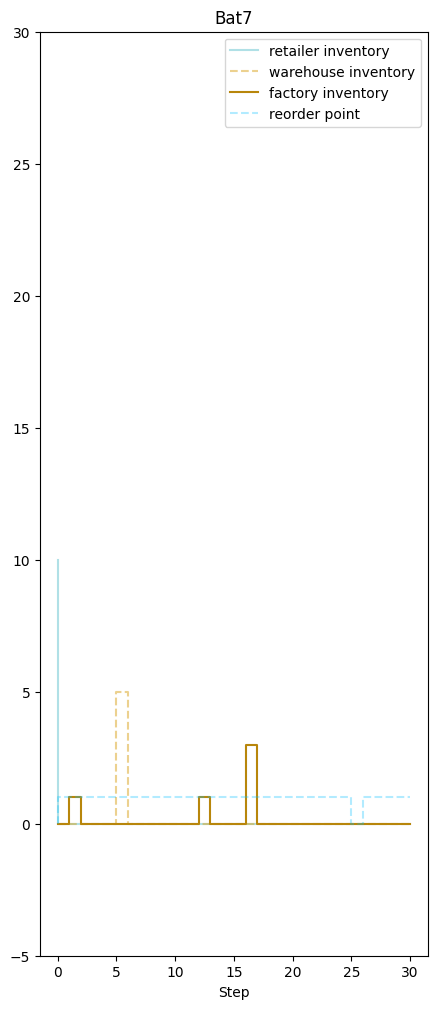}
\includegraphics[width=1in, height=0.34\linewidth]{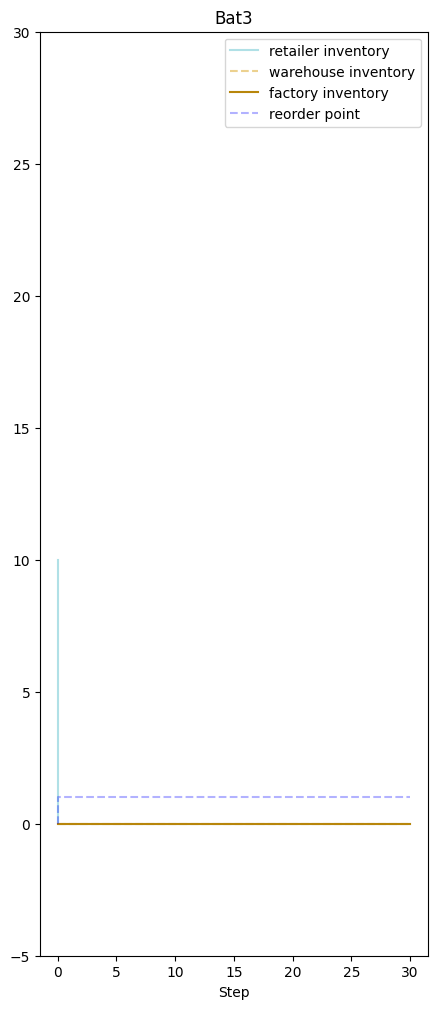}
  \quad
\caption{RPPO agent inventory resulting from the different strategies, which remain more consistent than in PPO}
    \label{fig:RPPO inventory policy }
\end{figure}

Figure \ref{fig:RPPO inventory policy } shows the resulting  inventory levels resulting from RPPO agent's different strategies. These stay almost identical to each other at 0.322 across all 6 environments. The same remains true for the warehouse and the factory inventories, which remain at an inventory level of 0.

Overall, the PPO agents don’t show much of a difference in outcome between the stochastic and batch environments, showing a similar reward level. There is a perceptible difference in strategies between them when looking at the actions which means that on some episodes it might be trying to control the stock while in others it accepts an early stockout. Each of the agents finds a loophole to reduce the episode length, as it finds controlling the system too challenging. The RPPO agents show a distinct behaviour between the least varying, Sto0 and batch, environments and the stochastic environments. However, they show a limited variation in behaviour between these environments and this has no real difference on the reward function. The agents avoid the loophole on the least varying environments, meaning they can control the performance at a similar level and indicates that these environments are easier to control. However, on the stochastic environments, the same loophole is discovered and used.

\section{Continuous reinforcement learning}
\label{Continual reinforcement learning}

In continuous reinforcement learning, the agent learns across different tasks, adapting to the varying tasks without having to retrain, reducing the cost as it doesn't require constant monitoring by a designer and where the performance is more stable. However, the ability for agents to learn across tasks has not been explored in semi-continuous non-stationary environments, similar to Supply Chain Environments. Experiments are performed across six environments from more to less variable. Initial tests are performed on the environments that are most similar to each other, the batch environments and then the stochastic environments. These are tested in both directions, from most variable to least and from least variable to most, to see if there is a curriculum learning benefit from going from the simplest problem to the most complex. 

\begin{figure*}[!htbp]
    \centering
  \includegraphics[width=7.2 in, height=0.27\linewidth]{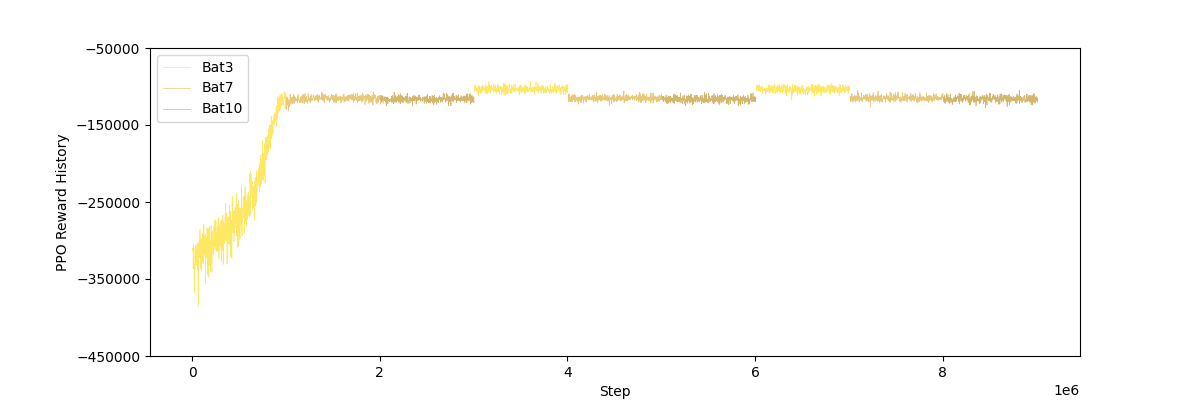}
  \quad
\includegraphics[width=7.2 in, height=0.27\linewidth]{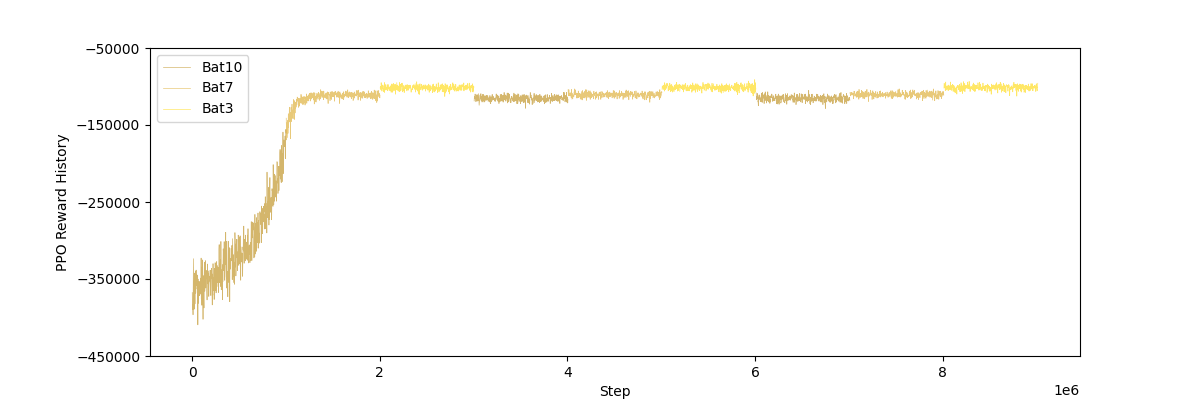}
\includegraphics[width=7.2 in, height=0.27\linewidth]{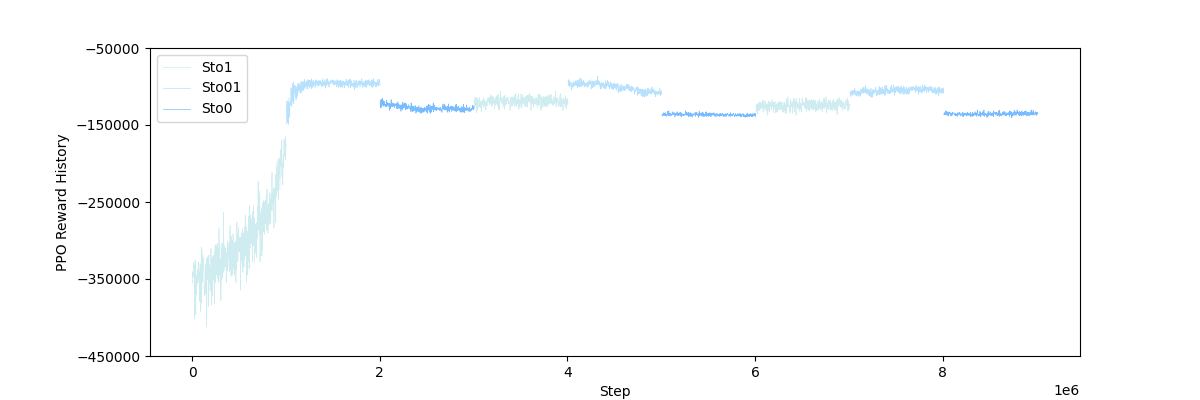}
\includegraphics[width=7.2 in, height=0.27\linewidth]{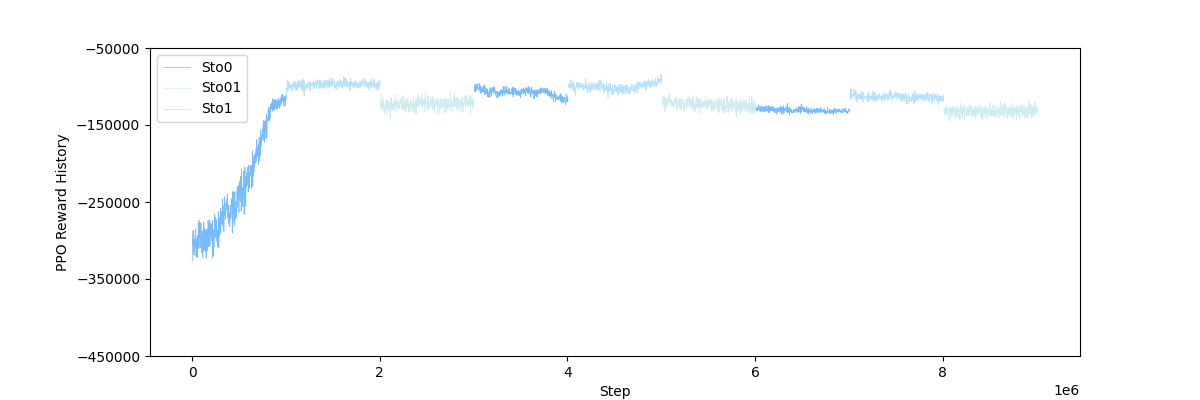}
    \caption{PPO agent's reward performance on the continuously varying supply chain with dark to light colours indicating the transition from less stochastic to more stochastic supply chain environments.}
    
    \label{fig:PPO's continuous}
\end{figure*}

Figure \ref{fig:PPO's continuous} shows results for the PPO agent. The figures show limited differences in performance between the environments. The agents all learn quickly in the first phase on all of the problems. The batch problems show a smaller difference in reward between the different tasks, with a slight increase in performacne for Batch 3 and the same maximum reward for batch 7 and batch 10. For the stochastic problems then there is a bigger range of differences between the different tasks with the Stochastic 0.1 environment having the greatest performance. However, the order of the tasks makes no difference. The agents are easily able to transfer their learning between the problems and avoid catastrophically forgetting tasks they have not seen recently. 

\begin{figure*}[!htbp]
    \centering
    \includegraphics[width=7.2in, height=0.27\linewidth]{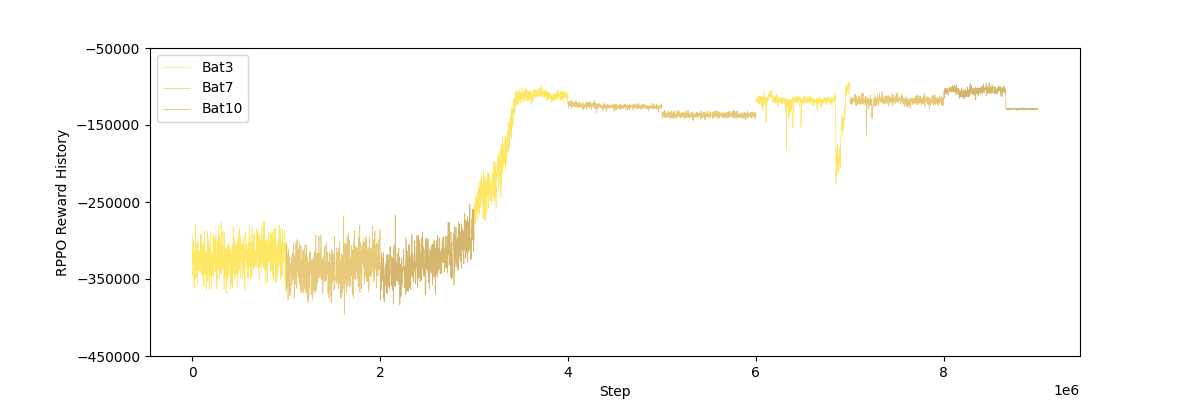}
    \quad 
    \includegraphics[width=7.2 in, height=0.27\linewidth]{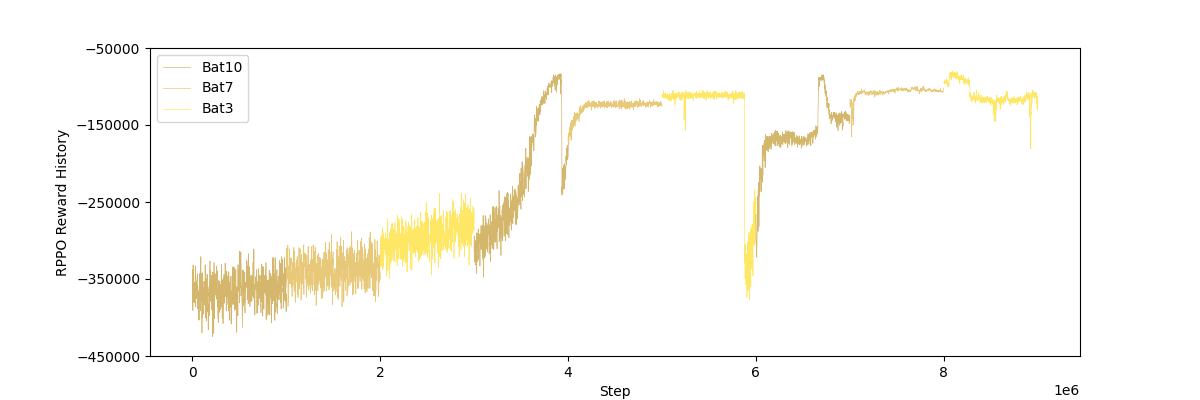} \includegraphics[width=7.2 in, height=0.27\linewidth]{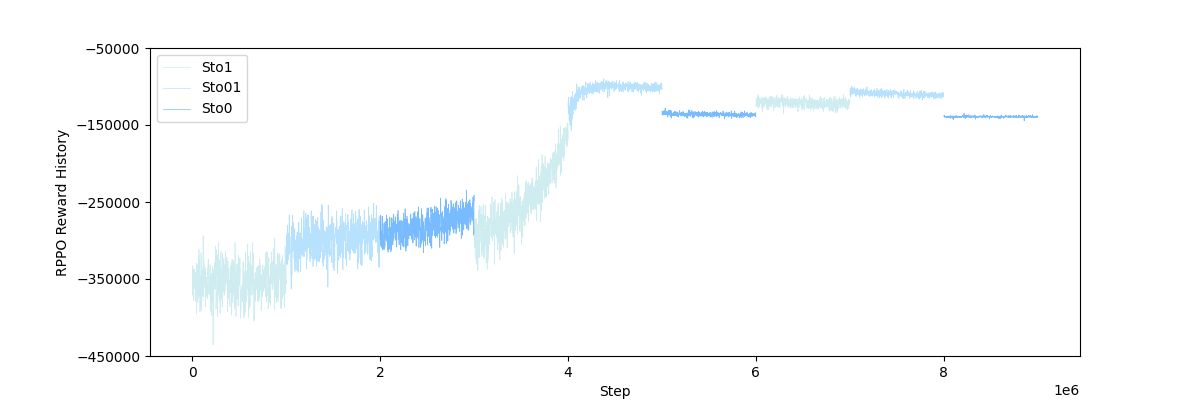}
\includegraphics[width=7.2 in, height=0.27\linewidth]{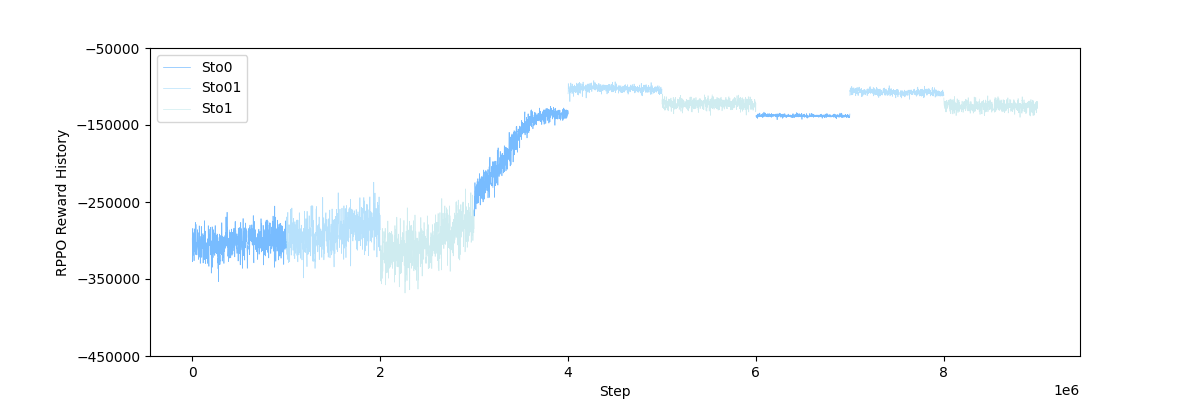}

    \caption{RPPO agent's reward performance on the continuously varying supply chain with dark to light colours indicating the transition from less stochastic to more stochastic supply chain environments.}
    \label{fig:RPPO's continuous}
\end{figure*}

For the RPPO agent the results for the continuous learning environment are shown in Figure \ref{fig:RPPO's continuous}. In this case the performance for the continuous learning is worse than for PPO. This is especially the case for the Batch environments, where the Recurrent element should give the biggest advantage. The agent struggles to transfer its learning between the different tasks and shows some drops in performance that are similar to those seen in \cite{sultana2020reinforcement}. It appears the agent is developing a specific strategy for each environment that doesn’t apply well to the other environments. However, for the stochastic environment the learner is able to more easily transfer it’s learning between tasks, giving a similar performance to PPO. Due to the lack of repeatability in the orders, the recurrent element doesn’t give an advantage and the agent performance is similar to that without the recurrent component, although the learning rate is lower and convergence happens later in the life. 

\begin{figure*}[!htbp]
    \centering
    
\includegraphics[width=7.2 in, height=0.27\linewidth]{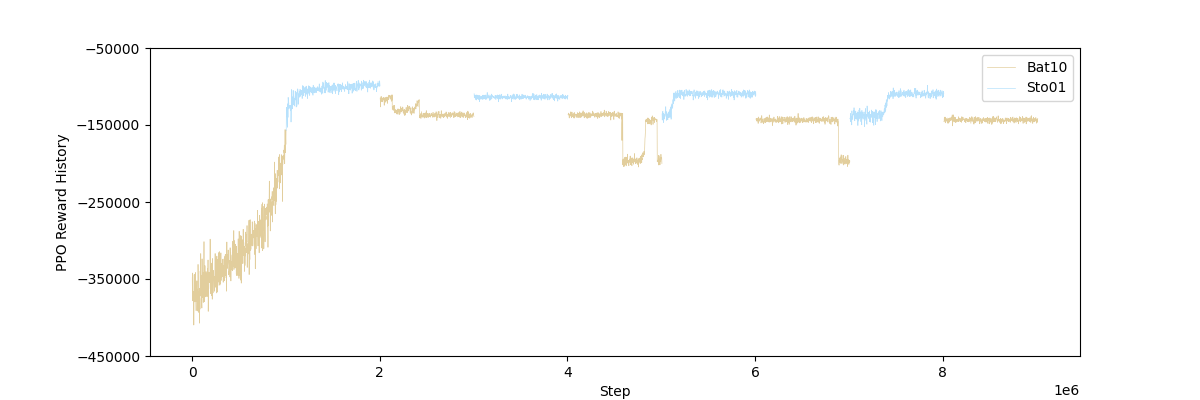}
\includegraphics[width=7.2 in, height=0.27\linewidth]{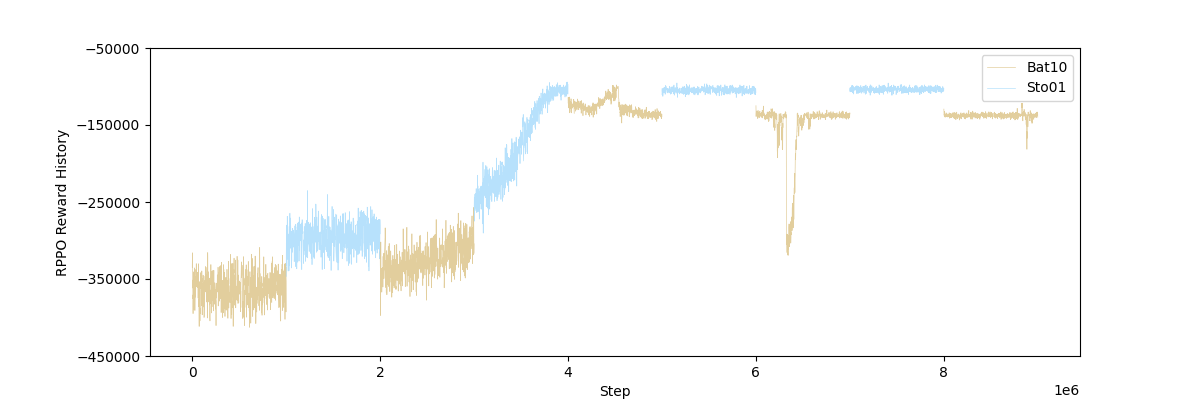}    \includegraphics[width=7.2 in, height=0.27\linewidth]{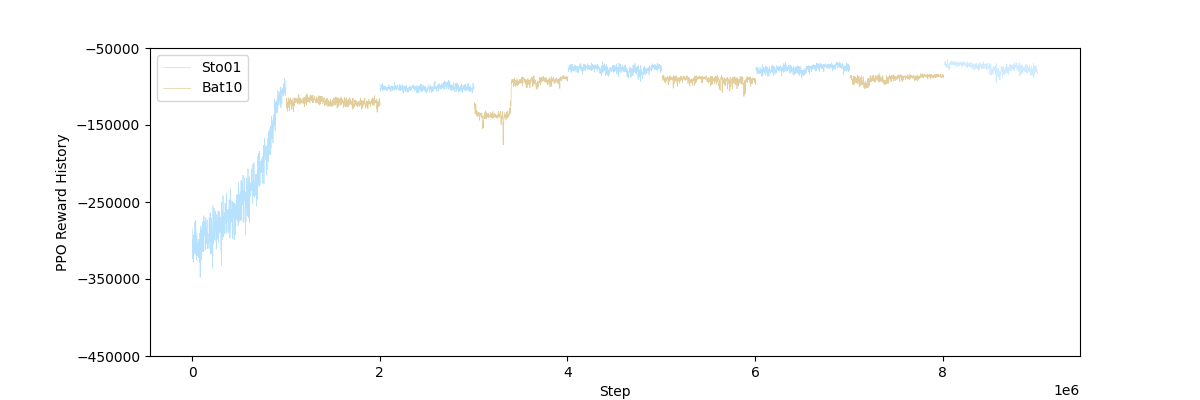}
\includegraphics[width=7.2 in, height=0.27\linewidth]{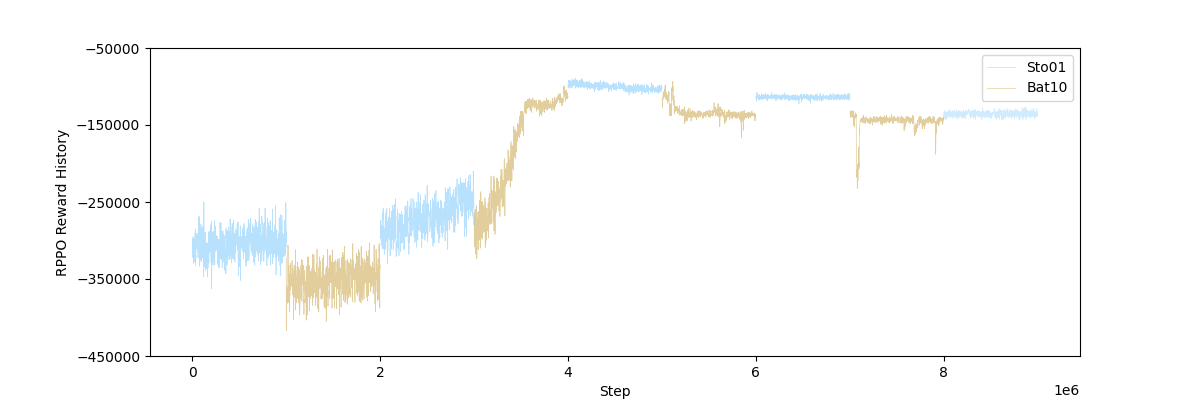}

    \caption{Reward curve from the PPO and RPPO agents continuously learning on supply chain environments with extreme variations.}
    \label{fig:bat to sto}
\end{figure*}

A more extreme change in environments is tested from the longest batches to the least varying of the stochastic orders and in reverse, with the results shown in Figure \ref{fig:bat to sto}. In these environments then the agents both show a poorer performance, with a larger jump in rewards between the environments and with dips in performance at a regular basis. The performance is better when going from the stochastic environment to the lean environment, showing that the more lean strategies the agents develop don't perform well on the more stochastic environments. The PPO agent still reaches a converged performance rapidly, with the RPPO taking a larger number of cycles to reach the same point. The RPPO agent in particular shows a regular dip in performance throughout the run and these dips are larger. The more extreme environments require differing strategies that are harder to transfer. This is supported by the results from the individual tests, where the strategies are different between the agents, but that this is hard to highlight purely from the reward curves.  

\section{Discussion and limitations}
Continuous Reinforcement learning is investigated in this paper through differences in the demand, this is because demand is the characteristic of the environment that is most likely to require a change in strategy. However, continuous learning could also be used when other aspects of the problem change, for example: inventory cost, service time, observation space and reward function. This is supported by previous research which shows that transfer learning on some of these problems was possible \cite{oroojlooyjadid2022deep} and the results from this paper indicate that continuous learning would be expected to be successful in these scenarios. This avoids having to train the parameters of the agent for each new problem, which considerably reduces the training time and means that human monitoring of the supply chain isn’t required. 

While the reward curves in the different scenarios show similar behaviours, there is a larger difference between the learners when the actions are investigated. It is common in the supply chain literature to compare the rewards alone, but the similarity could be a symptom of the initial reward structure. A larger variation in rewards between the retailer, warehouse and factory could provide a considerable difference in reward but it is difficult to say whether the behaviour itself is any more realistic. In this scenario it appears that the agents have learned to game the system, they have decided that it is easiest to take the stockout penalty rather than have a longer game where the penalty for having inventory leads to a larger overall penalty, especially in the environments where it is harder to control; those that are more stochastic or those that appear to vary every step as the agent has no memory. This is common in Reinforcement Learning where the optimal solution is often not found in more complex environments \citep{Bossens2019} as the initial search is random, providing agents an ability to learn came overcome these issues or through intrinsic rewards \citep{burda2018large}. 

Another approach could be a change in the environment through a more severe stockout penalty, through continuing the episodes no matter the number of stockouts, or by considering an average or reward that penalises shorter episode lengths. Another potential issue is the use of shared rewards, based on \cite{mousa2023analysis}, this ideally leads to more stable training but in this case does not penalise the factory agent making it harder to determine the root cause of the problem. No stock at the factory means that there is no ability to control the rest of the supply chain. However, in the batch environments, the RPPO, with an ability to remember histories, is able to overcome this to some extent and takes on a more realistic strategy. In other action sequences than the one highlighted a more positive strategy can also be seen, but the reward is still similar to the one highlighted in this paper. The environment used here is taken from the open literature and highlights the need for more carefully constructed environments for testing multi-agent models in dynamic supply chain environments as the findings are as related to the environments as the learners. 

Here the main findings are related to the change in behaviour in different environments, which are not invalidated by the agents using this loop hole. Since the PPO agent always finds the same strategy it transfer learns well, but the RPPO struggles to translate it's findings as the strategies are more distinct, despite more here still meaning that the strategies are quite similar. However, the agents still struggle on the continuous learning, meaning that even small differences in strategies can create these instabilities. This contradicts findings seen in toy problems like cart pole, where the similarity in task allows easy transfer learning \citep{bossens2021lifetime}, even this level of stocasticity in input creates a more challenging transfer environment. 

\section{Conclusion}
Continuous learning remains a problem for reinforcement learning, in many environments the agents fail to transfer learn and can catastrophically forget. In this paper, we demonstrate how Continuous Reinforcement Learning can be used for dynamic inventory control and investigate the behaviour of agents under these learning conditions. Experiments on six different task instances are reported, this adds a new type of demand to the common literature, where currently each product is likely to be different from the previous one in sequence. This new environment allows the products to stay the same for periods of time, assuming the decision maker is able to batch the orders. This new environment represents the opportunity for a wider range of lean strategies from the agent. Proximal Policy Optimisation (PPO) and Recurrent Proximal Policy Optimisation (RPPO) are tested on each of the 6 environments individually. Recurrent approaches are increasingly popular in the Supply Chain literature for online learning and would be expected to perform well on the more lean environments due to their memory of previous events. However, these approaches show a greater instability on all of the problems. The advantage of understanding what comes before is limited and leads to an over optimisation of the strategy in some cases meaning that it is more susceptible to small changes in demand. In the continuous learning environments PPO demonstrated similar strategies between environments and is easily able to transfer learn. However, it games the system as it is harder to find a realistic optimal strategy. This strategy is suboptimal but stable to the stochastic input and easier to find for the learner. The order for how the tasks appears doesn't change the performance in the continuous learning environment.  However, the RPPO agent struggles to transfer learn between the Batch environments, indicating that there is a strategy change between these different environments despite their similarity. When the environments are cycled between the most variable and least variable environments both the PPO and RPPO agents have periods where they perform poorly on the batch problems, this indicates that there is interference between these environments. However, it is not clear how realistic this large change in demand might be. Overall the paper demonstrates that continuous learning is possible, meaning that agents will need limited monitoring to control supply chains that vary over time.

\section*{ACKNOWLEDGMENTS}
The authors acknowledge the use of the IRIDIS High-Performance Computing Facility,
and associated support services at the University of Southampton, in the completion of this work. Wan Wang is funded by the China Scholarship Council. We would like to thank Lloyd's Register Foundation for their support during this research. Any errors or discrepancies are our own responsibility. The authors would also like to thank Wei Zhang for sharing his expertise in simulation optimisation with us during the preliminary stages of this research. 


\bibliography{references.bib}\label{refs}
\bibliographystyle{informs2014}

\end{document}